\RequirePackage{fix-cm}
\documentclass[twocolumn]{svjour3}          
\smartqed  
\usepackage{graphicx}
\usepackage[T1]{fontenc}
\usepackage[utf8]{inputenc}

\usepackage[usenames, dvipsnames]{xcolor}
\definecolor{linkcolor}{rgb}{0.0,0.3,0.5}
\usepackage[unicode,colorlinks=true,linkcolor=linkcolor,urlcolor=linkcolor,citecolor=linkcolor,pdfusetitle]{hyperref}
\usepackage{cite}
\usepackage[all]{hypcap}
\usepackage{graphicx}
\usepackage{xspace}
\usepackage{mathrsfs}
\usepackage{pifont} %
\usepackage{lmodern}

\expandafter\let\csname equation*\endcsname\relax
\expandafter\let\csname endequation*\endcsname\relax

\usepackage{amsmath,amsfonts,amssymb}
\usepackage{bm}
\usepackage{relsize, exscale}

\usepackage{longtable}

\usepackage{microtype}
\usepackage[normalem]{ulem} %

\usepackage[]{subcaption}

\renewcommand{\vec}[1]{\bm{\mathrm{#1}}}
\newcommand{\data}{\ensuremath{s}}
\newcommand{\vdata}{\ensuremath{\vec{\data}}}
\newcommand{\noise}{\ensuremath{n}}
\newcommand{\vnoise}{\ensuremath{\vec{\noise}}}
\newcommand{\signal}{\ensuremath{h}}
\newcommand{\vsignal}{\ensuremath{\vec{\signal}}}
\newcommand{\sigparams}{\ensuremath{\vec{\vartheta}}}
\newcommand{\covmat}{\ensuremath{\varSigma}}
\newcommand{\vcovmat}{\ensuremath{\vec{\Sigma}}}
\newcommand{\eigenvec}{\ensuremath{u}}

\newcommand{\eigenmat}{\ensuremath{U}}
\newcommand{\veigenmat}{\ensuremath{\vec{\eigenmat}}}

\newcommand{\vevalmat}{\ensuremath{\vec{\Lambda}}}
\newcommand{\acf}{\ensuremath{R}_{ss}}
\newcommand{\psd}{\ensuremath{S_n}}

\newcommand{\ip}[2]{\left< #1, #2 \right>}
\newcommand{\transpose}{\ensuremath{\mathsf{T}}}

\journalname{Neural Computing and Applications}
\begin{document}

\title{Genetic-algorithm-optimized neural networks for gravitational wave classification}


\author{Dwyer S. Deighan \and
Scott E. Field \and
Collin D. Capano \and
Gaurav Khanna}


\institute{Dwyer S. Deighan \at
          Department of Mathematics,
          Computer \& Information Science, 
          University of Massachusetts, 
          Dartmouth, MA 02747
          \email{ddeighan@umassd.edu}
          Dartmouth, MA 02747.
          \and
          Scott E. Field \at
          Department of Mathematics, 
          Center for Scientific Computing \& Visualization Research, 
          University of Massachusetts, 
          Dartmouth, MA 02747. 
          \email{sfield@umassd.edu}
          \and
          Collin D. Capano\at
          Max-Planck-Institut f\"ur Gravitationsphysik,
          Leibniz Universit{\"a}t Hannover,
          D-30167 Hannover, Germany
          \email{collin.capano@aei.mpg.de}
          \and
          Gaurav Khanna \at
          Department of Physics, 
          Center for Scientific Computing \& Visualization Research, 
          University of Massachusetts,
          Dartmouth, MA 02747. \& 
          Department of Physics,
          University of Rhode Island,
          Kingston, RI 02881.
          \email{gkhanna@umassd.edu}
}

\date{Received: date / Accepted: date}

\maketitle

\begin{abstract}
Gravitational-wave detection strategies are based on a signal analysis technique known as matched filtering. Despite the success of matched filtering, due to its computational cost, there has been recent interest in developing deep convolutional neural networks (CNNs) for signal detection. Designing these networks remains a challenge as most procedures adopt a trial and error strategy to set the hyperparameter values. We propose a new method for hyperparameter optimization based on genetic algorithms (GAs). We compare six different GA variants and explore different choices for the GA-optimized fitness score. We show that the GA can discover high-quality architectures when the initial hyperparameter seed values are far from a good solution as well as refining already good networks. For example, when starting from the architecture proposed by George and Huerta, the network optimized over the 20-dimensional hyperparameter space has 78\% fewer trainable parameters while obtaining an 11\% increase in accuracy for our test problem. Using genetic algorithm optimization to refine an existing network should be especially useful if the problem context (e.g. statistical properties of the noise, signal model, etc) changes and one needs to rebuild a network. In all of our experiments, we find the GA discovers significantly less complicated networks as compared to the seed network, suggesting it can be used to prune wasteful network structures. While we have restricted our attention to CNN classifiers, our GA hyperparameter optimization strategy can be applied within other machine learning settings.
\end{abstract}

%
%

\section{Introduction} \label{sec:intro}

During their first and second observing runs, the advanced Laser Interferometer Gravitational-Wave Observatory (LIGO) \cite{2015CQGra..32g4001L} and Virgo \cite{gw-detectors-Virgo-original-preferred} ground-based gravitational wave (GW) detectors have identified several coalescing compact binaries
\cite{DiscoveryPaper,LIGO-O1-BBH,2017PhRvL.118v1101A,LIGO-GW170814,LIGO-GW170608,LIGO-GW170817-bns,LIGO-O2-Catalog}. As these detectors improve their sensitivity, GW detections~\cite{Abbott:2016blz, TheLIGOScientific:2017qsa, Abbott:2016nmj,
Abbott:2017vtc, Abbott:2017gyy, Abbott:2017oio, LIGOScientific:2018mvr} are becoming routine~\cite{Aasi:2013wya, LIGOScientific:2018jsj}. 
In the current observing run, for example, gravitational wave events are now being detected multiple times a month~\cite{alerts}.
Among the most important sources for these detectors are binary black hole (BBH) systems, in which two black holes (BHs) radiate energy through GW emission, causing them to inspiral, merge, and finally settle down into a single black hole through a ringdown phase. GWs and their strong emission from compact astrophysical objects like binary black holes, are century-old predictions of Einstein's general relativity theory that have just recently been directly verified by the LIGO/Virgo network of detectors.

Current BBH gravitational wave search analysis~\cite{GWDA,LIGO-O2-Catalog} is based on a technique known as matched-filtering~\cite{1057571}. The detector's output, i.e. time-series data of the relative motion of the mirrors as a function of time, is correlated (i.e. ``matched'') with a set of expected signals known as templates. These templates are generated using theory-based models of expected GW sources. To find all signals buried in a given dataset, a complete catalog of templates should cover all astrophysically plausible signals we might expect to observe. Consequently, templates must sample the BBH parameter space with sufficient density, which results in very large catalogs and computationally expensive analysis~\cite{Harry:2016ijz}. There are currently several low-latency pipeline configurations~\cite{gstlal,spiir,cWB,mbta2016,pyCBClive} that partially reduce this expense by a combination of hardware acceleration and algorithm-specific optimizations. 

It is known that the matched filter is the optimal linear filter for maximizing the chances of signal detection in the presence of an additive, Gaussian noise. Yet despite its remarkable successes, the main drawback of matched-filtering is its high computational expense. Furthermore, this optimality result is limited by two very strong assumptions, Gaussian noise and knowing the expected signal precisely. Clearly, these assumptions are not satisfied in practice, and so modern search efforts have extended the simple matched-filter framework to work in realistic settings~\cite{LIGO-O2-Catalog}.

Deep filtering~\cite{George2018FirstPaper} is an alternative, machine-learning-based approach that has received significant attention over the past two years~\cite{shen2019deep,hezaveh2017fast,levasseur2017uncertainties,ciuca2019convolutional,gabbard2018matching,shen2017denoising,george2017glitch,george2018deep,fort2017towards,George2018FirstPaper,gebhard2019convolutional} as a way to overcome the aforementioned limitations of matched filtering. While it remains to be seen if deep filtering can entirely replace matched filtering in realistic settings, it is primarily due to its orders-of-magnitude faster performance that makes it a very promising candidate to complement traditional search pipelines in the context of low-latency detection. As a first step towards this goal, multiple researchers have demonstrated that deep filtering can achieve accuracy comparable to that of matched filters~\cite{gabbard2018matching}. There has already been significant exploration of different approaches to deep filtering, involving recurrent neural networks~\cite{shen1711denoising}, transfer learning~\cite{george2018classification}, topological feature extraction~\cite{bresten2019detection}, Bayesian networks~\cite{lin2020detection}, binary neutron stars~\cite{krastev2020real,schafer2020detection,lin2020binary}, and multiple detectors~\cite{fan2019applying}. In the context of gravitational-wave data analysis, deep learning has been shown to be highly-effective for low-latency signal searches and parameter estimation~\cite{george2018deep,chua2020learning,gabbard2019bayesian,green2020gravitational}, with and without Gaussian noise, detector glitch classification~\cite{george2017glitch}, denoising gravitational waves~\cite{shen2017denoising,wei2020}, and even to characterize the GW signal manifold~\cite{khan2020}. 

While there has been significant attention paid to different approaches to deep filtering, one aspect of the problem that has gone unexplored, 
however, is an automated approach to hyperparameter optimization. This issue arises in two scenarios. First, when testing out a new deep 
filtering network on an entirely new class of signals or noise. In such cases, it is unknown what the hyperparameter values should be and a brute force search is often used.  Given the large number of hyperparameters, typically around 20 for the cases we will consider, a human tester might wish to test 10 different values for each hyperparameter resulting in an unacceptably large $\approx 10^{20}$ different network configurations to train. Hyperparameter optimization may also be needed to refine an existing network. For example, perhaps an already good network architecture is known, but this network was trained on a specific signal class and noise model. If, say, the noise characteristics (due to non-stationary drifts) or the signal model changes, one might be interested in finding optimal network configurations using an already reasonable network architecture as a starting guess. To maximize a deep filter's search sensitivity, it makes sense to invest extra offline computational resources to identify an improved network.

Despite the importance of hyperparameter optimization, and numerical optimization being a well-studied subject, to our knowledge there are no currently agreed-upon best practices to accomplish this. Indeed, this is an open area of inquiry taking place in many different fields of research and industry~\cite{ul2014optimization,goodfellow2016deep,AWSage,hamdia2020efficient}. Within the gravitational wave community, the only 
methods considered have been brute force searches by trial-and-error.

In this paper, we develop a class of genetic algorithms (GAs) to solve this problem. GAs are optimization algorithms seeking to (in our case) maximize a fitness function. GAs are built from a collection of random walkers exploring the fitness function along with evolution-inspired heuristic rules for these walkers to interact, mutate, and move towards regions of high fitness. Briefly, the algorithm begins with a random population of network architectures, then iterates through 3 phases: selection, crossover, and mutation. The selection phase occurs when the candidates compete for survival based on their fitness. The crossover phase is the first search operator, where some of the surviving candidates swap their hyperparameter values (called genes in GA literature) and replace themselves. The mutation phase is the second search operator, where the genes may undergo a random walk to a nearby value. These concepts will be explored more fully, but similarities with particle swarm optimization, which have been recently explored in gravitational-wave contexts~\cite{pso_gw}, are noteworthy.

We will show that GAs can automatically discover new Deep Filtering networks that are both more accurate and more compact, which will allow searches with these refined networks to detect more signals more quickly. We also provide comparisons between GA variants and Monte Carlo. As GW detectors are exceptionally sensitive to very massive objects \cite{AstroPaper}, and the majority of compact binaries observed to date are pairs of ${\cal O}(30M_\odot)$  BBH systems \cite{LIGO-O2-Catalog}, we will focus on such systems. 

This paper is organized as follows. In Section~\ref{sec:preliminaries} we introduce the GW detection problem and signal detection diagnostics. 
In Section~\ref{sec:deep_filter} we summarize the deep convolutional neural network filter and the hyperparameters that define its architecture. This architecture is optimized for a fitness score using a family of related GAs that are introduced in Sec.~\ref{sec:GA}. Numerical experiments on a variety of benchmark tests are considered in Sec.~\ref{sec:numerical_experiments}. Our experiments focus on exploring the properties of the genetic algorithm and its performance under different scenarios.

\section{Preliminaries} \label{sec:preliminaries}

This section summarizes the gravitational-wave signal detection problem, which provides a framework for answering the question:
is there a gravitational-wave signal in the data? We review this background material to facilitate a clearer context for the
convolutional neural networks considered throughout this paper.

Signal detection in general, and gravitational wave detection in particular, is a well established field. This section primarily draws
from~\cite{Maggiore2008,Owen:1995tm,DBrownThesis,Cutler:1994ys,romano2017detection,Wainstein:1962,Allen:2005fk}, and our conventions follow those of Ref.~\cite{Allen:2005fk}.

\subsection{Gravitational-wave signal model}
\label{sec:GW_model}

A gravitational-wave strain signal $h(t)$ detected by a ground-based interferometer has the form,
\begin{align}
\label{eq:gw}
h(t;\sigparams) = &\frac{1}{r}F_{+}\left(\text{ra},\text{dec},\psi\right) h_{+}(t;\iota,\phi_c,t_c,M,q) + \nonumber \\
                     &\frac{1}{r}F_{\times}\left(\text{ra},\text{dec},\psi\right) h_{\times}(t;\phi_c,t_c,,M,q) \, ,
\end{align}
where $r$ is the distance from the detector to the source, $t_c$ is the coalescence time (time-of-arrival),  $\phi_c$ is the
azimuthal angle between the $x-$axis of the source frame and the line-of-sight to the detector (sometimes called the orbital phase at coalescence), $\iota$ is the inclination angle between the orbital angular momentum of the binary and line-of-sight to the detector, and the antenna patterns $F_{(+,\times)}$ project the gravitational wave's $+$- and $\times$-polarization states, $h_{(+,\times)}$, into the detector's
frame. The antenna patterns are simple trigonometric functions of variables which specify the orientation of the detector with respect to the binary: the right ascension ($\text{ra}$), declination ($\text{dec}$), and polarization ($\psi$) angles. For the non-eccentric, non-spinning BBH systems considered here are typically parametrized by a mass ratio $q = m_1/m_2 \geq 1$ and total mass $M = m_1 + m_2$, where $m_1$ and $m_2$
are the component masses of each individual black hole. To summarize, the measured gravitational-wave strain signal $h(t;\sigparams)$ is described by $9$ parameters, $\sigparams = \{r, \phi_c, \iota, \text{ra}, \text{dec}, \psi, t_c , M, q \}$ whose range of values will be set later on.

When discussing waveform models, it is common practice to introduce the complex gravitational wave strain
\begin{align}
h_{+}(t;\iota, \phi_c, \dots) &- {\mathrm i} h_{\times}(t;\iota, \phi_c, \dots) \nonumber \\
& =  \sum_{\ell=2}^{\infty} \sum_{m=-\ell}^{\ell} h^{\ell m}(t;\dots) {}_{-2}Y_{\ell m} (\iota, \phi_c) \, ,
\end{align}
which can be decomposed~\cite{NewmanPenrose1966,Goldberg1967} into a complete basis of spin-weighted spherical harmonics ${}_{-2}Y_{\ell m}$. Here, for brevity, we only show the model's dependence on $\iota$ and $\phi_c$ while suppressing the other $7$ parameters.
Most gravitational waveform models make predictions for the modes, $h^{\ell m}$, from which a model
of what a noise-free detector records, $h(t;\sigparams)$, is readily recovered. 

Throughout this paper we will consider a numerical relativity gravitational-wave surrogate model that provides up to $\ell \leq 8$ harmonic modes
and is valid for $1 \leq q \leq 10$~\cite{Blackman:2015pia}. We evaluate the model through the Python package GWSurrogate~\cite{gwsurrogate,Field:2013cfa}.

\subsection{Signal detection problem setup}
\label{sec:signal_detection_setup}

Consider a single gravitational-wave detector. We sample the output of the detector at a rate $1/\Delta t$ over some time period $T$. This produces a set of $N = T/\Delta t$ time-ordered samples $\vdata$.\footnote{For simplicity, we assume here that $N$ is even. This can always be made to be the case, since the observation time and sampling rate are free parameters in an analysis.} In the absence of a signal, the detector is a stochastic processes that continually outputs random noise $\vnoise$.  We wish to know whether a gravitational-wave signal $\vsignal$ exists in the detector during the observation time, or if the detector data consist purely of noise.  This is complicated by the fact that the signal depends on the unknown value of $\sigparams$. For now, we simplify the problem by asking whether the data contains a signal with fixed parameters $\sigparams'$ (we will relax this condition later). In that case, our problem is reduced to finding a statistical test that best differentiates between two simple hypotheses, the signal hypothesis $H_1': \vdata = \vsignal(\sigparams') + \vnoise$ and the null/noise hypothesis $H_0: \vdata = \vnoise$.

Let $\beta$ be the probability of making a type II error with our test (the false dismissal probability), so that $1-\beta$ is its power (the
probability that we reject the noise hypothesis when the signal hypothesis is true), and $\alpha$ be the probability of making a type I error with our test (the false alarm probability). By the Neyman-Pearson lemma \cite{Neyman:1933wgr}, the most powerful test that can be performed between two simple hypotheses at a significance level $\alpha$ is the likelihood-ratio test.  That is, given the likelihood
ratio,
\begin{equation}
\Lambda(\vdata|\sigparams') = \frac{p(\vdata|\sigparams', h)}{p(\vdata|n)},
\end{equation}
we reject the noise hypothesis if $\Lambda(\vdata|\sigparams')$ exceeds a threshold value. Here the vertical bar denotes a conditional probability. For example, $p(\vdata|\sigparams', h)$ is the probability of observing the signal, $\vdata$, given a gravitational waveform model $h$ and system parameters $\sigparams'$.

To proceed further, we need to assume a model for the noise. It is standard to assume that the detector outputs wide-sense stationary Gaussian noise such that the Fourier coefficients of the noise, $\tilde{n}(f_i)$, satisfy
\begin{align} \label{eq:noise_fd}
\langle \tilde{n}(f_i) \rangle = 0 \,, \qquad \langle \tilde{n}(f_i) \tilde{n}^*(f_j) \rangle = \frac{T}{2} S_n(f_i) \delta_{ij}  \,,
\end{align}
where the brackets, $\langle X \rangle$, denote the expectation value of a random variable $X$, $S_n(f)$ is the single-sided power spectral density (PSD) computed from $n(t)$, and $\delta_{ij}$ denotes the Kronecker delta function. In this case, the likelihood (see Appendix~\ref{app:matched_filter} for a derivation) that the data does not contain a signal is 
\begin{equation}
\label{eqn:noise_likelihood}
p(\vdata | \noise) \propto \exp\left[-\frac{1}{2} \ip{\vdata}{\vdata}\right] \,,
\end{equation}
and we do not need to evaluate the normalization constant as it will cancel in the likelihood ratio. The inner product is defined as
\begin{equation}
\label{eqn:discreteip}
\ip{\vec{a}}{\vec{b}} \equiv 4 \Re \left\{\Delta f \sum_{p=p_0}^{N/2-1} \frac{\tilde{a}^{*}[p]\tilde{b}[p]}{\psd[p]} \right\} \,,
\end{equation}
where $*$ denotes complex conjugation, $\Delta f = 1/T$, $\psd$[p] is the PSD of the noise evaluated at frequency $f =p \Delta f$,
$\tilde{a}[p]$ indicates the Fourier transform of the time domain vector $\mathbf{a}$ evaluated at frequency $f = p \Delta f$, and $p_0$ corresponds to a low frequency cutoff, below which the PSD is effectively infinity; for current generation detectors, this is at $\sim 20\,$Hz.

Since the signal hypothesis is $\vdata = \vsignal(\sigparams') + \vnoise$, the likelihood that the data contains a signal is simply the probability of observing $\vnoise = \vdata - \vsignal(\sigparams')$,
\begin{equation*}
p(\vdata | \sigparams', \signal)
    \propto \exp\left[-\frac{1}{2} \ip{\vdata - \vsignal(\sigparams')}{\vdata - \vsignal(\sigparams')}\right] \,,
\end{equation*}
assuming the same noise model. The likelihood ratio is therefore
\begin{equation}
\label{eqn:likelihood_ratio}
\Lambda(\vdata|\sigparams') = \exp\left[\ip{\vsignal(\sigparams')}{\vdata} - \frac{1}{2} \ip{\vsignal(\sigparams')}{\vsignal(\sigparams')}\right].
\end{equation}
Since this only depends on the data via the $\ip{\vsignal(\sigparams')}{\vdata}$ term, a sufficient statistic for the simple hypothesis test is
\begin{equation}
\label{eqn:simple_stat}
K = \ip{\vsignal(\sigparams')}{\vdata}.
\end{equation}
Note that in the literature $K$ is often taken to be $K = | \ip{\vsignal(\sigparams')}{\vdata} |$ to account for large, negative values that indicate that the data contains the signal, but that it is $180^\circ$ out of phase with the test signal. As we will see later, $K$ is related 
to the signal's SNR whose statistical properties, in turn, depend  on this choice. In particular, in the absence of a signal the test statistic Eq.~\eqref{eqn:simple_stat} is normally distributed with zero mean and variance $\sigma^2 = \ip{\vsignal(\sigparams')}{\vsignal(\sigparams')}$
\cite{DBrownThesis}. With the alternative choice, $K$ would have been $\chi$-distributed with one degree of freedom.

To indicate whether or not there is a signal in the data, we can use the one-sided test function,
\begin{equation}
\label{eqn:simple_test}
\varphi'(\vdata) =
    \begin{cases}
    1 & \text{if } \ip{\vsignal(\sigparams')}{\vdata} \geq K*, \\
    0 & \text{otherwise} \,,
    \end{cases}
\end{equation}
with the threshold $K^*$ chosen such that the size of the test is
\begin{equation}
\label{eqn:simple_size}
\mathbb{E}_{H_0} \varphi'(\vdata) = \int_{\ip{\vsignal(\sigparams')}{\vdata} \geq K^*} \exp\left[-\frac{1}{2}\ip{\vdata}{\vdata}\right] \mathrm{d}\vdata \leq \alpha.
\end{equation}

As stated above, Eq.~\eqref{eqn:simple_test} is the most powerful test assuming fixed parameters. However, in practice, the parameters of the signal are not known a priori. We therefore need a test that can distinguish between the null hypothesis $H_0$ and a \emph{composite} hypothesis $H_1: \vdata = \vsignal(\sigparams) + \vnoise$, where the parameters $\sigparams$ may be in a range of possible values. 

\subsection{Detecting a signal with unknown amplitude}

Most of the signal parameters --- such as mass, spin, etc. --- cannot be analytically maximized over, as the signal models have non-trivial dependence on them. We can, however, construct a uniformly most powerful test that maximizes over the distance, $r$, since $1/r$ is simply an overall amplitude scaling factor for the signal~\eqref{eq:gw}.

To construct the optimal statistic that allows for any distance, consider a template signal $\vsignal$ that is generated at some fiducial distance
$r_0 > 0$ such that all possible astrophysical signals $\vsignal'$ are at a distance $r'\geq r_0$. The signal hypothesis becomes $H_1: \vdata =
A\vsignal(\sigparams) + \vnoise$, where $A \equiv 1/r \in (0, 1]$. Assume for a moment that we use the same test statistic and function as defined
in Eqs.~\eqref{eqn:simple_stat} and \eqref{eqn:simple_test}, but with $\vsignal_m(\sigparams')$ replaced with $A\vsignal_m(\sigparams)$. The statistical power is
\begin{align*}
& 1 - \beta \equiv \mathbb{E}_{H_1}\varphi'(\vdata) = \\
      & \int_{A \ip{\vsignal(\sigparams)}{\vdata} \geq K^*} \exp\left[A\ip{\vsignal}{\vdata} - \frac{1}{2} A^2\ip{\vsignal}{\vsignal} -\frac{1}{2}\ip{\vdata}{\vdata}\right] \mathrm{d}\vdata
\end{align*}
(for simplicity of notation, from here on we will use $\vsignal$ to mean $\vsignal(\sigparams)$). Since $A \in (0, 1]$, the power grows monotonically with $A$. Maximizing the argument, which noting that in the exponent over $A$ yields
\begin{equation}
\label{eqn:maxA}
A = \frac{\ip{\vsignal}{\vdata}}{\ip{\vsignal}{\vsignal}}.
\end{equation}
This gives test statistic $K = \ip{\vsignal}{\vdata}^2/\ip{\vsignal}{\vsignal}$, or, equivalently,
\begin{equation}
\label{eqn:maxamp_stat}
\rho = \frac{\ip{\vsignal}{\vdata}}{\sqrt{\ip{\vsignal}{\vsignal}}}
=  \ip{\hat{\vsignal}}{\vdata} \,,
\end{equation}
where we have defined a normalized template, $\hat{\vsignal}$, that satisfies $\ip{\hat{\vsignal}}{\hat{\vsignal}} =1$. The quantity $\rho$ is known as the \emph{signal-to-noise ratio} (SNR). Let $\vdata = C \hat{\vsignal} + \vnoise$, where $C \geq 0$ ($C=0$ corresponds to 
the noise hypothesis), then $\rho$ is normally distributed with the following mean and variance: 
\begin{align} \label{eq:SNR_properties}
\langle \rho \rangle = C \,, \qquad \langle \rho^2 \rangle - \langle \rho \rangle^2  = 1  \,.
\end{align}
To indicate whether or not there is a signal in the data, we can use the one-sided test function,
\begin{equation}
\label{eqn:maxamp_test}
\varphi(\vdata) =
    \begin{cases}
    1 & \text{if } \rho(\vdata) \geq \rho^*, \\
    0 & \text{otherwise,}
    \end{cases}
\end{equation}
where the threshold, $\rho^*$, is chosen such that the size $\mathbb{E}_{H0} \varphi(\vdata) \leq \alpha$. Note that $\varphi(\vdata)$ has the same size and power as $\varphi'(\vdata)$ for the simple hypothesis test $H_1'$ in which some fixed $A$ is used. This is because the two functions only differ by the factor of $1/\sqrt{\ip{\vsignal}{\vsignal}}$, which for the simple signal hypothesis is a constant. Consequently, $\varphi$ is the uniformly most powerful test for any distance $ r> r_0$. In terms of the SNR, the matched filtering classifier Eq.~\eqref{eqn:maxamp_test} will generate false alarms with a probability of
\begin{align} \label{eq:false_alarm}
\alpha(\rho_*) = p(\rho > \rho_* \rvert H_0 ) = \int_{\rho_*}^{\infty} p(\rho \rvert H_0 ) d \rho\,,
\end{align}
and false dismissals with a probability of
\begin{align} \label{eq:false_dismiss}
\beta(\rho_*) = p(\rho < \rho_* \rvert H_1' ) =  \int_{- \infty}^{\rho_*} p(\rho \rvert  H_1' ) d \rho \,.
\end{align}

\subsection{Matched-filter classification}

In practice one will need to search over the entire model space. We select a discrete set $\{\sigparams_i\}_{i=1}^M$ of $M$ parameter values and a corresponding {\em template bank} of normalized filters
\begin{align*}
{\cal B} = \{\hat{\vsignal}(\sigparams_i) \text{ s.t. } \sigparams_i
\in \{\sigparams_i\}_{i=1}^M \text{ and } \langle \hat{\vsignal}, \hat{\vsignal} \rangle = 1\} \,.
\end{align*}
The bank's SNR is defined to be
\begin{align} \label{eq:bank_snr}
\rho({\cal B}) = \max_i \rho \left( \sigparams_i \right) \,,
\end{align}
where $\rho \left( \sigparams_i \right)$ is the SNR computed with $\hat{\vsignal}(\sigparams_i)$. While each $\rho \left( \sigparams_i \right)$ is 
normally-distributed the bank's SNR, $\rho({\cal B})$, is not. The bank's efficacy will depend on how densely we sample the continuum.
A faithful template bank guarantees that for any possible signal with an optimal SNR (signal and filter are identical) of $\rho_\mathrm{opt}$, then one of the templates in the bank will be sufficiently close to the optimal one such that $\rho({\cal B}) \gtrapprox 0.97 \rho_\mathrm{opt}$.

To summarize, assuming a Gaussian noise model, the matched-filter signal-detection classifier is to test if the bank SNR is larger than a predetermined threshold value. Sec.~\ref{sec:deep_filter} will described the convolutional neural network signal-detection classifier for solving the same signal-detection problem.

\subsection{Signal detection diagnostics}
\label{sec:diagnostics}

One goal of this paper is to compare different CNN-based classifiers. The diagnostics we will use to facilitate this comparison include the false alarm and dismissal probabilities, accuracy, receiver operating characteristic (ROC) curves, and efficiency curves.

\subsubsection{False alarms, dismissals, and accuracy}

Given a classifier ranking statistic, $R$, (for the matched filter this is the SNR, $R=\rho$, and for the CNN this is the output of the softmax layer, $R =P_{\tt signal}$) and a threshold, $R^*$, on this value, we can assign labels to our data. We can then compare the true labels to compute the number of false alarms and false dismissals. For certain cases, the false alarm, $\alpha$, and false dismissal, $\beta$, probabilities can be computed analytically. However, in many cases, in particular, for CNN classifiers, these probabilities can only be computed empirically through a Monte Carlo study. A true alarm is $1 - \beta$ while the true dismissal is $1 - \alpha$. These four numbers, $1 - \alpha$, $1 - \beta$, $\alpha$, and $\beta$, define the {\em confusion matrix}. In Sec.~\ref{sec:ga_fitness}, one component of the GA fitness score is the accuracy, which for a balanced testing set containing an equal number of examples with and without a GW signal, the accuracy is given by $1 - \alpha / 2 - \beta / 2$.

\subsubsection{Receiver operating characteristic}

An ROC curve plots the true alarm probability, $1-\beta(R^*)$, vs the false alarm probability, $\alpha(R^*)$, both of which are functions of the ranking statistic threshold $R^*$. Such curves can be used to assess the classification strategy as the detection threshold is varied. It is important to note that the shape of an ROC curve will depend on the anticipated distribution of the ranking statistic over a class of expected signals. For example, we expect different ROC curves for weak and strong signal strengths.

\subsubsection{Efficiency curves}

An efficiency curve plots the true alarm probability vs signal strength at a fixed value of either the ranking statistic threshold or false alarm probability. Such curves can be used to assess the classification strategy as the signal's power is varied. For very loud signals (SNRs> 15) we find that all CNN classifiers are essentially perfect, while for weaker signals (SNRs between 3 and 10) the classifier's efficacy will depend on details such as the architecture and problem domain. 

\section{Deep models for time series classification}
\label{sec:deep_filter}

Sec.~\ref{sec:preliminaries} summarized a classical matched-filtering approach to signal detection: given time series data $\vdata$ and a template bank of possible signals we compute the SNR~\eqref{eq:bank_snr} whose value provides both a classification method (exceeding a threshold) as well as a measure of significance. In this section, we summarize one commonly explored CNN that has been successfully used for the same purpose of signal detection. Our key aim will be to describe what parameters describe the CNN and their interpretation since it will be the genetic algorithm's job to optimize their values. 

Deep networks are specified by learned parameters and hyperparameters. Learned parameters are found through an optimization procedure known as training. Hyperparameters are parameters that control the network's overall design and characteristics, and unlike learned parameters, their values are provided by the programmer. We will distinguish between three flavors of hyperparameter. We will refer to the parameters used to describe the network's structure as {\em model hyperparameters}, and the ones we consider are summarized 
in Sections~\ref{sec:conv_layer} and \ref{sec:full_layer}. Those parameters that control the training process will be referred to as {\em training hyperparameters} are summarized in Sec.~\ref{sec:training_hparams}. Finally, since we have control over our training set we will consider {\em training-set hyperparameters}, summarized in Sec.~\ref{sec:training_set_hparams} to be those parameters that control the training set generation.

Usually, it's not clear what values the hyperparameters should be, so one must resort to trial-and-error or random sampling of the hyperparameter space. The main goal of our work is to automate the process of exploring the hyperparameter space with a genetic algorithm, introduced in Sec.~\ref{sec:GA}, such that the resulting network's architecture is optimized.

\begin{figure*}[thb]
\includegraphics[width=0.9\textwidth]{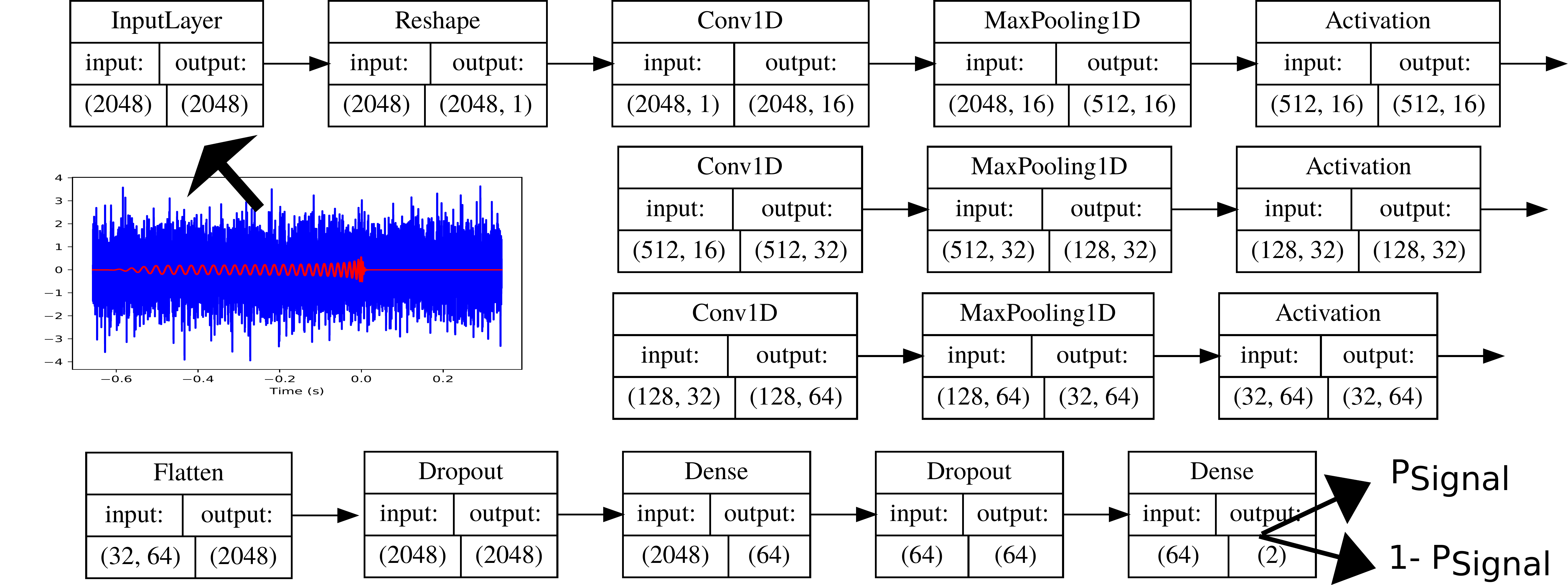}
\caption{Architecture of the typical classifier network used to seed the genetic algorithm optimizer. The input vector is 1 second of data sampled at 2,048 Hertz and the classification output layer uses the softmax activation function such that the vector's components sum to one. We interpret the value of $P_{\tt signal}$ as a measure of the network's confidence that the data contains a signal.  Three convolutional layers are used to extract signal features, which are subsequently passed through two dense layers each of which is composed of a fully-connected linear layer and a non-linear ReLU activation function. Dropout layers are used to reduce overfitting. Each layer has its own internal structures and parameters that are discussed in the text and summarized in Table.~\ref{tab:final_results}. The genetic algorithm will modify this seed architecture and each layer automatically by approximately solving the hyperparameter optimization problem. 
}
\label{Fig:classifier_network}
\end{figure*}

\subsection{Classifier network and its hyperparameters}
\label{sec:classifer_network}

Fig.~\ref{Fig:classifier_network} summarizes a typical classifier network considered in this paper, which is based on the original Deep Filter discovered by George and Huerta~\cite{George2018FirstPaper}. In fact, the overall architecture displayed in Fig.~\ref{Fig:classifier_network} is the same as their network except for the inclusion of two extra dropout layers that we use to reduce overfitting. The authors of Ref.~\cite{George2018FirstPaper} tested around 80 different network architectures, and for the best one(s) the network's hyperparameters were tuned manually via a trial-and-error procedure ~\cite{George2018FirstPaper}. 

From Fig.~\ref{Fig:classifier_network} we see that the input is first reshaped to match the input expected by a Keras' Conv1D layer. This is a trivial step that we mention only for completeness. Next, a sequence of convolutional layers is applied. In keeping with common terminology, we will refer to a single convolutional layer as built out of three more primitive layers: Conv1D, MaxPooling1D, and Activation, all of which are summarized in Sec.~\ref{sec:conv_layer}. From Fig.~\ref{Fig:classifier_network} we see that the initial input vector passes through three convolutional layers, after which it has been transformed into a matrix. The Flatten layer simply ``unwraps'' the matrix into a vector, which is subsequently passed through a sequence of two fully-connected layers. We will refer to a single fully-connected layer as built out of Dropout layer and a Dense layer, each of which is summarized in Sec.~\ref{sec:full_layer}. 

The output of the final layer is a vector with two components which sum to 1. The ranking statistic, $0 \leq P_{\tt signal} \leq 1$, is a measure of the network's confidence that the data contains a signal. Similar to the matched filtering case outlined in Sec.~\ref{sec:signal_detection_setup}, we can use the one-sided test function
\begin{equation}
\label{eqn:simple_test_cnn}
\varphi_{\tt CNN}(\vdata) =
    \begin{cases}
    1 & \text{if } P_{\tt signal} \geq P^*, \\
    0 & \text{otherwise.}
    \end{cases}
\end{equation}
to indicate whether or not there is a signal in the data.
The threshold $P^*$ can be chosen such that the size of the test
satisfies $\mathbb{E}_{H_0} \varphi_{\tt CNN}(\vdata) \leq \alpha$. 

The basic architecture structure enforced on our classifier network is a sequence of $N_{\tt conv}$ alternating convolutional and pooling layers followed by $N_{\tt full}$ fully-connected layers. Currently, the number of layers is set manually although in principle the genetic algorithm could be extended to modify these values allowing it to explore shallower or deeper networks. As described in Sec.~\ref{sec:GA}, we penalize deepness through an overall network complexity factor that modifies the fitness score~\eqref{eq:GA_fitness}.

\subsubsection{The convolutional layer's hyperparameters}
\label{sec:conv_layer}

The first part of the network is described by $N_{\tt conv}$ convolutional layers. This part of the network seeks to extract learned features from the data which are local and may appear at different locations of the dataset. A single convolutional layer is typically built out of three more primitive sub-layers~\cite{goodfellow2016deep}, which is depicted in Fig.~\ref{Fig:classifier_network} as Conv1D, MaxPooling1D, and Activation.

The Conv1D layer performs a discrete convolution of the input data with $C_{\tt filter}$ kernels, which are sometimes called filters. The convolution's output is designed to extract features from the signal, and so if there are $C_{\tt filter}$ filters our output data will provide information about the possible appearance of $C_{\tt filter}$ features. Filters are typically compact with a window length (or size) specified by $C_{\tt filter-size}$ and in any given Conv1D layer all filters are required to have the same size. The convolution can be modified by specifying a stride factor, $C_{\tt filter-stride}$, the number of data samples to move per convolution operation. Finally, we consider dilated convolutions that effectively enlarge the filter by a dilation rate, $C_{\tt filter-dilation}$, to explore larger features of the data.

The next sub-layer is MaxPooling1D, which is a max reduction over an input vector of size $P_{\tt size}$. The pooling operator is slid across the input vector spaced $P_{\tt stride}$ elements apart. For example, in Fig.~\ref{Fig:classifier_network} we see that the MaxPooling1D operation applied in the first convolutional layer takes the input vector of size $2048$ to an output vector of size $512$ ($=2048 / 4$), which means $P_{\tt stride} = 4$. As the max reduction is applied to each filter's output, the input and output vector's row size is left unchanged.

The final sub-layer uses the common Rectified Linear Unit (ReLU) activation function that simply requires the output is positive by applying a function, $\max(x,0)$, to each element of the input vector. This layer has no hyperparameters and so does not contribute to the search space.

To summarize, the $i^\mathrm{th}$ convolutional layer is uniquely defined by 6 numbers $C^i_{\tt filter}$, $C^i_{\tt filter-size}$, $C^i_{\tt filter-stride}$, $C^i_{\tt filter-dilation}$, $P^i_{\tt size}$, and $P^i_{\tt stride}$. We consider $N_{\tt conv}$ convolutional layers and allow different hyperparameter values in each layer. So in total there are as many as $6 N_{\tt conv}$ 
hyperparameters associated with the network's convolutional layers.

\subsubsection{The fully-connected layer's hyperparameters}
\label{sec:full_layer}

The second part of the network is described by $N_{\tt full}$ fully-connected layers. As we will always use dropout, we will refer to a fully-connected neural network layer as built out of two more primitive sub-layers, which is depicted in Fig.~\ref{Fig:classifier_network} as Dropout and Dense.

Input to the first densely-connected layer is a set of features provided by the last convolutional layer. The goal of the densely-connected layers is to find a non-linear function mapping the features to the correct classification signal vs no-signal. That this might even be possible, in principle, one often appeals to the universal approximation theorem~\cite{hornik1989multilayer,cybenko1989approximation}. However, neither this theorem nor any we are aware of, provide guidance on the number or depth of the layer that should be used for a particular problem.

The Dropout sub-layer randomly sets a random fraction, $D_{\tt drop}$, of the input units to zero at each training update. As such, the network after dropout can be viewed as a smaller layer (fewer neurons) that is forced to train on the dataset same. This technique helps to reduce overfitting. There are no learned weights in this sub-layer.

The final Dense sub-layer is a neural network connecting all of the inputs to $D_{\tt units}$ output units. We use a ReLU activation function for all fully-connected layers except the final one. The final output layer's activation is the softmax function, which maps a real number to the interval $[0,1]$.

To summarize, the $i^\mathrm{th}$ fully-connected layer is uniquely defined by $D^i_{\tt drop}$ and $D^i_{\tt units}$. We consider $N_{\tt full}$ fully-connected layers and allow different hyperparameter values in each layer. The $i^\mathrm{th}$ fully-connected layer is uniquely defined by $D^i_{\tt drop}$ and $D^i_{\tt units}$. So in total there are as many as $2 N_{\tt full}$ 
hyperparameters associated with the network's fully-connected layers.

\subsubsection{Training hyperparameters}
\label{sec:training_hparams}

Given some value for the model and training-set hyperparameters we seek to learn good values for the weights by solving an optimization problem seeking to minimize a loss function. Training hyperparameters affect the solution's convergence properties and computational resources.

We use the well-known ADAM optimizer~\cite{kingma2014adam} to solve this optimization problem. ADAM works by estimating the gradient on a subset of the training data known as the batch size, $N_{\tt batch}$. This optimizer has three hyperparameters, a learning rate, $\epsilon_{\tt LR}$, and two adaptive moment decay rates, $\beta_{\tt Adam1}$ and $\beta_{\tt Adam2}$. The optimizer will continue until either reaching a maximum number of iterations (or epochs), $N_{\tt epochs}$, or the validation error steps decreasing for a predetermined number 
of iterations. In all of our experiments, we use the standard categorical cross entropy loss function.

In some numerical experiments, we allow the GA to modify a subset of training hyperparameters over a restricted range. In some cases, like with the number of epochs, the values are set mostly by considering the computational cost. For other cases, as with adaptive moment decay rates, good default values are known and so requiring the GA to explore the enlarged dimensionality of the hyperparameter space is likely not worthwhile. We note that the ADAM optimizer already exploits automatic modification of the learning rate that changes with the iteration.

\begin{table*}[]
\centering
\caption{Hyperparameters that determine the classifier network. These parameters may control the overall network's architecture or properties of an individual layer. A network is uniquely specified (up to its learned weights) by architecture and layer parameter values. The learning of the network's weights, which are found by solving an optimization problem, are controlled by the training parameters. The optimal network's weights, in turn, implicitly depend on the training set parameters. Some parameter values are fixed to reduce the dimensionality search space, in which case we quote typical values used in our experiments. For GA-modified parameters, the Valid Range column denotes the largest range the GA could explore (sometimes called the prior in Bayesian optimization). However, in practice, the population of hyperparameter solutions explore regions localized around the seed network (cf.~ Sec.~\ref{sec:hp_intervals} and Fig.~\ref{fig:comp_spread_converg}).
}

\label{tab:hparams}
\centering
\begin{tabular}{|c|c|c|c|c|}
\hline\hline
Parameter & Description & Type & GA Modifies & Valid Range \\ \hline \hline

\multicolumn{5}{|c|}{Model hyperparameters} \\ \hline \hline


$N_{\tt conv}$ & \# of 
Conv1D layers & Architecture & No & \{3,4,5\}\\
$N_{\tt full}$ & \# of 
dense layers & Architecture & No & \{2,3\} \\
$C^i_{\tt filter}$ & Number of filters & $i^\mathrm{th}$ Conv1D layer & Yes &  [1,600] \\
$C^i_{\tt filter-size}$ & Filter size & $i^\mathrm{th}$ Conv1D layer & Yes & [1,600] \\
$C^i_{\tt filter-stride}$ & Filter stride & $i^\mathrm{th}$ Conv1D layer & Yes & [1,600] \\
$C^i_{\tt filter-dilation}$ & Filter dilation & $i^\mathrm{th}$ Conv1D layer & Yes & [1,600] \\
$P^i_{\tt size}$ & Pooling size & $i^\mathrm{th}$ Pooling layer & No & 4\\
$P^i_{\tt stride}$ & Pooling stride & $i^\mathrm{th}$ Pooling layer & No & 4\\
$D^i_{\tt drop}$ & Dropout rate & $i^\mathrm{th}$ Dropout layer & Yes & [0,0.5] \\
$D^i_{\tt units}$ & Output units & $i^\mathrm{th}$ Dense layer & Yes & [1,600]\\
\hline \hline
\multicolumn{5}{|c|}{Training hyperparameters} \\ \hline \hline
$N_{\tt batch}$ & Batch size & Adam Optimizer & Yes & [32, 64] \\
$\epsilon_{\tt LR}$ & Learning Rate & Adam Optimizer & Yes & [$10^{-5}$, $10^{-3}$] \\
$\beta_{\tt Adam1}$ & Moment decay & Adam Optimizer & Yes & [0.8, 0.999] \\
$\beta_{\tt Adam2}$ & Moment decay & Adam Optimizer & Yes & [0.95, 0.999999] \\
$N_{\tt epochs}$ & Epochs & Training & No & [80, 600] \\
$N_{\tt patience}$ & Patience & Training & No & 8 \\
\hline \hline
\multicolumn{5}{|c|}{Training-set hyperparameters} \\ \hline \hline
$N_{\tt TS}$ & Training examples & Training set & No & [10, $10^{4}$] \\
$f_{\tt signal}$ & Fraction of signals & Training set & No & [0 , 1] \\
 \hline
\end{tabular}
\end{table*}

\subsection{The training set and its hyperparameters}
\label{sec:training_set_hparams}

When preparing training data we can control the overall number of training examples, $N_{\tt TS}$, and the fraction of training examples containing a signal, $f_{\tt signal}$. We consider these training-set hyperparameters as they are not learned yet control the final classifier model. Ideally, we would like to $N_{\tt TS}$ as large as possible, however larger training sets can lead to much longer training times and can be excessive in some cases. Indeed, we have found that for loud signals (say, SNRs greater than 100) perfect classifier networks can be trained with just tens of training examples while many thousands of examples are needed for weak signals at low SNRs. For now we have not allowed the GA to modify training-set hyperparameters.

We use a training strategy inspired by George and Heurata's technique of presenting the classifier network with training data of increasing
difficulty by decreasing the SNR~\cite{George2018FirstPaper}. They found that this strategy was able to improve the classifier's final
accuracy while reducing the overall training time. We decrease the SNR by increasing noise amplitude rather than manipulating the distance parameter, and our target SNR is the average SNR over the dataset, where individual signals will have SNR values distributed around the average. 
In addition to decreasing the SNR, we simultaneously increase the parameter domain's extent by slowly widening an initially narrow sampling 
distribution around the target parameter interval's mean to the full interval. The full problem is thus revealed to the network over a 
specified number of datasets until the parameter intervals and SNR reach their largest extent and smallest value, respectively. We provided example values in the numerical experiments section.

The typical sizes of time-domain gravitational-wave data are of the order $10^{-21}$. With such small values, it is well-known that deep networks require the training data to be normalized to train correctly.  A common choice is to whiten the data by the PSD, such that after whitening each training example has a zero mean and unit variance. We have pursued a PSD-agnostic approach whereby a normalization layer is the first network layer (not shown in Fig.~\ref{Fig:classifier_network}) that is used to achieve a target mean absolute deviation (MAD) of the input signal. For example, if we set our target MAD value to be 1000, and the training data's MAD is $10^{-19}$, we would multiply the input data by $10^{22}$. Through trial and error we found a target MAD of $1000$ to work well for our problem, although we also explored letting the GA optimize for this value. We also tried batch normalization before the input layer but it appeared to not work as well.

\begin{figure}[thb]
\centering
\includegraphics[width=0.5\textwidth]{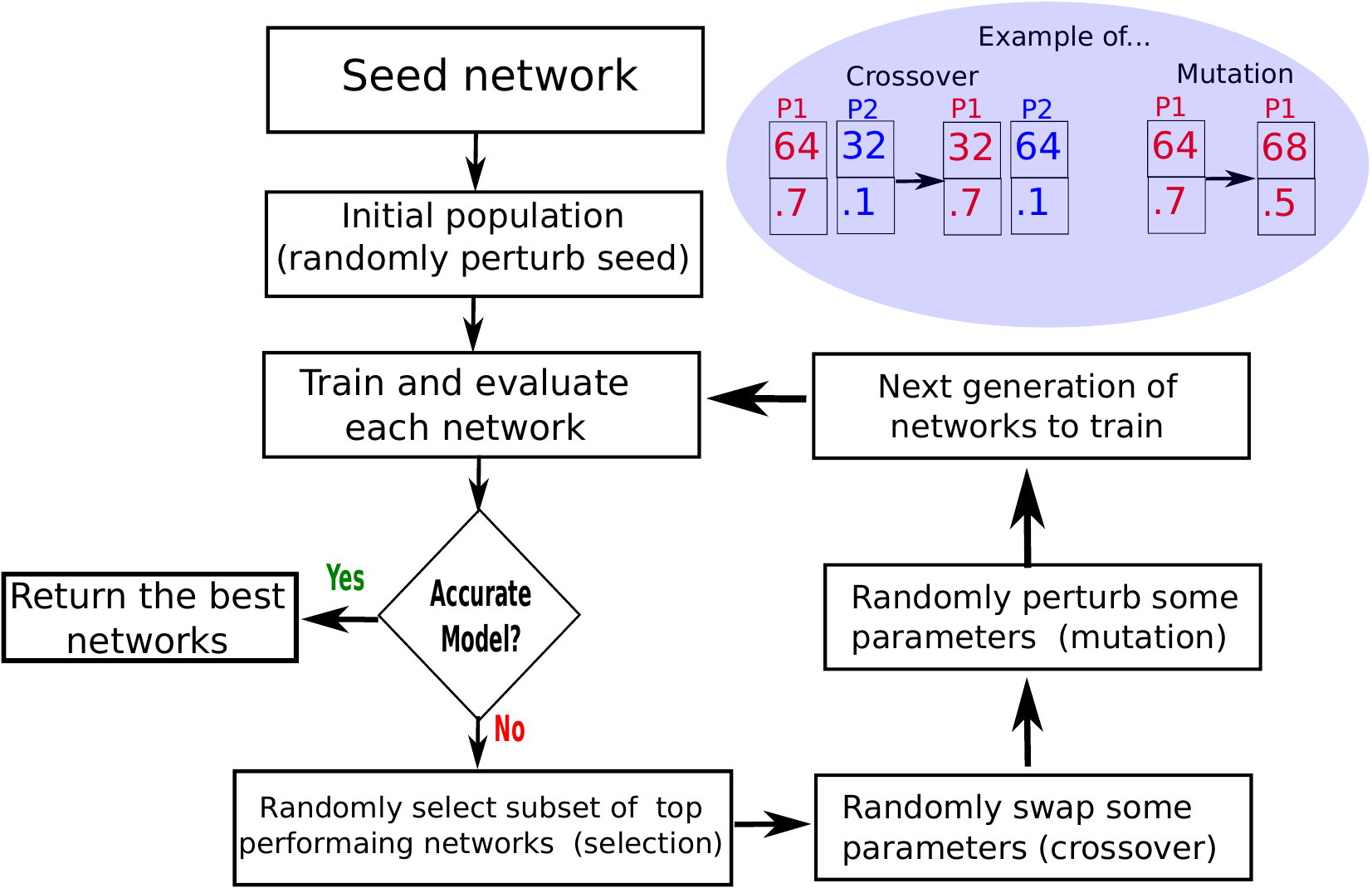}
\caption{Genetic algorithm hyperparameter optimization. The algorithm starts with an initial population
 of networks created by randomly sampling a large volume of the hyperparameter space centered around the seed network. Each network is trained and an overall fitness score~\eqref{eq:GA_fitness} is computed. A sequence of selection, crossover, and mutation operators are applied to the population, thereby generated a new population to train. This process continues until either a suitably accurate network is identified or the populations have converged to a best solution. The overall performance of the optimization process requires that we use reasonable values for the search operators. We explore six different choices in this paper. The inset bubble shows the effect of 1-point crossover and mutation operators acting on candidate models.}
\label{Fig:GA}
\end{figure}

\subsection{An optimization model for the hyperparameters}
\label{sec:optimization_model}

Table~\ref{tab:hparams} summarizes the various hyperparameters that will impact the final trained classifier network. Regardless of the algorithm used to solve the hyperparameter optimization problem, it is helpful to know in advance what parameters should be improved, how they should be changed, and any constraints or relationships that should be enforced between them. While there is not a general theory applicable to our problem, our choices are guided by insights compiled by previous efforts to design similar classifiers~\cite{George2018FirstPaper,shen2019deep,gebhard2019convolutional,gabbard2018matching} as well as our own expectations. 

For example, some parameters should not be modified by the GA. The training size ($N_{\tt TS}$), epochs ($N_{\tt epochs}$), and early-stopping condition ($N_{\tt patience}$), for example, are problem-specific numbers that can be set by available computational resources and common sense. In our case, $N_{\tt TS}$ and $N_{\tt epochs}$ is often set as large as possible such that training a single network can be completed in under 24 hours. We also do not allow the optimizer to change the number of convolutional or dense layers, which would dramatically alter the network's behavior; finding for good values of $N_{\tt conv}$ or $N_{\tt full}$ are better accomplished through a simple grid-based search while optimizing over the remaining set of modifiable hyperparameters. While the pooling-layer parameters could be modified for some problems, in our case we do not. Indeed, the pooling stride is essentially redundant with the convolutional layer stride (which we optimize for). The pooling size is often used to add some
network robustness to time translations of the input vector, so that signals shifted in time will still be detected. Since the intended use of our signal-detection classifier is for continuously streamed data~\cite{George2018FirstPaper,gebhard2019convolutional}, and the signal's time of arrival is of astrophysical importance, we prefer the network to be sensitive to time translations. We do use a small, fixed value of the pooling size for downsampling the data and reducing training time.

Training hyperparameters have a well-known range of good values that have been extensively used in the literature. This fact is reflected in the tight intervals shown in Table~\ref{tab:hparams}. Plausible values for the moment decay parameters, for example, are similar to the ranges suggested in the original ADAM optimizer paper~\cite{kingma2014adam}. For different deep networks, various scaling relations connecting 
$N_{\tt batch}$ and $\epsilon_{\tt LR}$ have been proposed, typically of the form $\epsilon_{\tt LR} \propto N_{\tt batch}$ or $\epsilon_{\tt LR} \propto \sqrt{N_{\tt batch}}$ ~\cite{hoffer2017train,smith2017don,goyal2017accurate}. While such relations could be used to reduce the problem's dimensionality, we have not explored this possibility here. Our numerical experiments (cf. Fig.~\ref{fig:optimizer_2D_grid_LR_BS}) suggest the existence of a similar scaling relationship for our problem.  

The plausible range and relationship between the remaining model hyperparameters is somewhat less clear and will be highly problem dependent. For example, while it is known that changing the stride and dilation of the convolutional filter will detect signal features of different characteristic sizes, we do not know ahead of time what these values should be for gravitational wave signals embedded in detector noise. Lacking a trustworthy model for these parameters, we allow the optimization procedure to fully explore this hyperparameter subspace over a relatively large region. In all of our GA experiments, the final optimized hyperparameter values do tend to lie within a factor of $\approx 3$ from the starting values of the George \& Huerta seed network. Yet unconstrained optimization does result in some surprises: contrary to our expectation based on other CNN architectures reported in the literature, the middle convolutional layer typically has the fewest number of filters after being optimized by the GA (cf. Table.~\ref{tab:final_results}).

\section{Hyperparameter optimization with genetic algorithms}
\label{sec:GA}

\subsection{Motivation}

Hyperparameter optimization is difficult. For example, Table~\ref{tab:hparams} lists at least $18$ hyperparameters defining the classier model with potentially many more as certain hyperparameters are defined layer-by-layer, and so the total number of hyperparameters will grow with the number of layers. Furthermore, the gradient of the relevant objective function either may not exist (e.g. discrete-valued parameters) or may be noisy, and we might expect to train hundreds or thousands of networks, so the algorithm must parallelize efficiently. 

There are not many optimization algorithms that meet the above conditions. Due to the dimensionality of the problem, a brute-force grid search would be prohibitive while gradient-based optimization is unavailable due to the formal lack of a gradient. Consequently, in gravitational-wave applications, the hyperparameters have been selected by a combination of experience, intuition, and random sampling. While the resulting networks have been accurate, they are not expected to be optimal. Nor is it known how close to the optimal configuration they might be. Their architectures might be biased by intuition or unnecessarily complicated for a given problem. 

Evolutionary algorithms are a class of optimization algorithms that meet all of the above criteria. We consider one particular variant of an evolutionary algorithm for hyperparameter optimization known as genetic algorithms~\cite{back2018evolutionary}. These algorithms have been inspired by concepts of natural evolution and survival of the fittest. They are stochastic optimizers drawing on familiar ideas.

Genetic algorithms due come with some drawbacks which include they have their own hyperparameters to set (fortunately setting these parameters is relatively easy) and they can require significant computational resources to evaluate many candidate models in parallel. As with any optimization algorithm, its possible they will get stuck in local minima. Since the optimization of the hyperparameters is an offline cost it is reasonable to use all available computational resources to search for the best network configuration. In our case, many of our numerical experiments took just a few days using 20 compute nodes with NVIDIA Tesla V100 GPUs. To avoid local minima, a few independent GA simulations can be performed or the mutation rate can be increased.  

We first summarize the essential pieces that make up a genetic algorithm then, later on, provide specific configurations considered and compared throughout this paper.

\subsection{General algorithmic workflow}

The algorithm's structure is summarized in Fig.~\ref{Fig:GA}. One complete iteration of the inner-loop is called a {\em generation}, and this process continues for multiple iterations or until a sufficiently small value of the model's fitness score is found. A list of the top models are recorded throughout all generations, and their hyperparameter values and scores are returned when the algorithm is finished.

The algorithm begins with a seed value for the hyperparameter, ${\bf \lambda}_{\mathrm seed}$, where ${\bf \lambda}$ is a vector of model hyperparameters. In the GA literature this vector is sometimes called a {\em chromosome} and its components are known as a {\em genes}. Starting from a seed, a set of, say, 20 candidate hyperparameter 
values, $\{{\bf \lambda}_{i}^1\}_{i=1}^{20}$, are drawn from a probability distribution as described in Sections~\ref{sec:mutation} and~\ref{sec:hp_intervals}. Here the notation ${\bf \lambda}_{i}^j$ means the hyperparameter values of the $i^\mathrm{th}$ candidate model for the $j^\mathrm{th}$ iteration of the genetic algorithm. At any given iteration the set of all surviving solutions is called the {\em population}.

Next, each candidate classifier model (defined by its value ${\bf \lambda}_{i}$) is trained, validated, and an overall fitness score is computed. The fitness score may attempt to maximize accuracy, minimize architecture complexity, penalize false positives, or any other desirable property. In particular, it need not be the loss function used for training. Our particular choice (cf. Sec.~\ref{sec:ga_fitness}) defines the fitness score as a weighted sum of the validation accuracy and an estimate of the model's complexity. In this way, the population of classifier models will be nudged towards simpler models.

A key aspect of any genetic algorithm is to continually update the population so that it moves toward higher values of the fitness score. This is achieved by applying a sequence of three operators to the population $\{{\bf \lambda}_{i}^1\}_{i=1}^{20}$. These are referred to as a {\em selection operator}, {\em crossover operator}, and finally a {\em mutation operator}. Taken together, these three operators will generate new candidate hyperparameter values sometimes referred to as children, which are subsequently added to the population. 

Next, we will describe in more detail these three operators as they are defined for one particular variant of the $(\mu + \lambda)$-evolutionary algorithm, which in turn is one particular class of genetic algorithms we will consider. 

\subsubsection{Selection operator}

Our first step in this process is to select a subset of top-performing models. The best two can be automatically selected (known as the elitism selection rule, which we will sometimes use), while from the remaining models we randomly pair off in subsets of 2 and select the best one of the subset. This procedure, known as tournament-style (or arena) selection rules, continues until we are left with $\mu$ models in total. Tournament selection is performed with replacement and can be generalized to have subsets of more than 2 competing for selection. Note that our selection rule does not simply pick the best $\mu$ models, but rather randomly selects $\mu$ models that are biased towards the fittest while inferior solutions are removed with a higher probability.

The remaining $\mu$ models function as parents. The parent model seed $\lambda$ new models (known as children) according to a set of operations described below. Consequently, after this step, there will be $\lambda$ children models and $\mu$ parent models. The $(\mu + \lambda)$-evolutionary algorithm allows both parent and children models to continue to the next iteration giving a population size of $\lambda + \mu$ candidate models. Despite the increased population size, there are only $\mu$ new models to train. In a variant strategy, which we will refer to as the ``standard'' evolutionary algorithm, only the $\lambda$ children models will be part of the next generation.

\subsubsection{Crossover operator}

We pair off randomly selected candidate models and swap their hyperparameter values with some probability known as the crossover rate, $p_{\tt cross}$. This is sometimes referred to as breeding in the GA literature. Two popular options are the one-point and two-point crossover operators. Each randomly selects position(s) in the hyperparameter vector where two solutions' content is spliced into each other. This operation allows for the generation of new candidate models. An example of a one-point crossover is shown in Fig.~\ref{Fig:GA}. Our genetic algorithms use both 1-point and 2-point crossover rules. Note that the order in which hyperparameters are stacked will impact the solution after crossover. We group hyperparameters that describe larger units together, which preserve higher-level structures. For example, parameters that specify each convolutional layer are grouped together in the hyperparameter vector.

\subsubsection{Mutation operator} \label{sec:mutation}

After crossover there is the mutation phase, this is where randomly selected solutions undergo mutation on randomly selected genes. We associate with each model some probability of changing its hyperparameter values known as the mutation rate, $p_{\tt mutate}$. If its selected for mutation, we then associate with each gene some probability of changing its value known as the gene-mutation rate, $p_{\tt gene}$. For the experiments used in this paper, we typically set $p_{\tt gene} = 1/N_{\tt CNN}$, where $N_{\tt CNN}$ denotes the dimensionality of the hyperparameter search space. We tried larger values of $p_{\tt gene}$ but they performed worst on the problems we considered. If a hyperparameter is selected for mutation its value is modified according to a Gaussian mutation: we draw the new value from a normal distribution whose mean is the current value and whose variance is $0.2 \times \left(I_m^\mathrm{high} - I_m^\mathrm{low}\right)$, where the parameter-specific interval $I_m$ is defined in Eq.~\eqref{eq:mapped_interval} and $I_m^\mathrm{high}$ and $I_m^\mathrm{low}$ are the upper and lower boundaries of this interval, respectively. 

\subsubsection{Fitness evaluation}
\label{sec:ga_fitness}

At the end of the modification steps we have a new set of $\mu$ candidate hyperparameter values, $\{{\bf \lambda}_{i}^{j}\}_{i=1}^{\mu}$, to add to the population. Each new candidate classifier model is trained, validated, and an overall fitness score is computed. The fitness score can be flexibly selected to encourage networks with desirable properties and, in particular, need not be related to the loss function used for training the network. We choose our fitness score to be
\begin{align} \label{eq:GA_fitness}
S_i^j = 0.975 J_{i}^j + 0.025 C_{i}^j \,,
\end{align}
where $J_{i}^j$ is the $i^\mathrm{th}$ classifier's accuracy at generation $j$, evaluated on the validation dataset,
and $C_{i}^j$ is network's size (or complexity) fitness. The weighting factors are selected such that $S_i^j \leq 1$.

The accuracy is computed using a simple formula as the number of correctly classified examples divided by the total number of examples. To assign a label to each testing example, we use a threshold of $P^* = 0.5$ in our one-sided test function~\eqref{eqn:simple_test_cnn}; we will return to the choice of threshold in Sec.~\ref{sec:network_comparisons}. The complexity fitness is computed as the ratio of the total number of trainable parameters (the network's degrees of freedom) computed relative to the seed network. As an example, $C_{i}^j$ obtains a maximum value of $1$ if there are no learned parameters, is $0$ if there are as many learned parameters as the seed network, and can be negative if there are more learned parameters than the seed network.  Note that there are many possible alternative measures for complexity one could consider, such as the Rademacher complexity or Vapnik-Chervonenkis dimension. In all cases, the complexity fitness measure should result in more compact networks (hence faster training times) and might lead to better performing networks by reducing generalization error or adversarial attack examples~\cite{yin2019rademacher}. 

The genetic algorithm's goal is to maximize the fitness score, which is a weighted sum of the accuracy and an estimate of the model's complexity. Table~\ref{tab:hparams} provides typical ranges we allow our hyperparameters to vary over.

\subsection{Genetic algorithm variants}
\label{sec:GA_variants}

To summarize, a completely specified genetic algorithm will specify a selection, crossover, and mutation operator. We mainly consider the following 6 variations in this paper:
\begin{itemize}
\item {\bf Standard}: The selection rule does not use elitism and only children comprise the next generation. We use a 1-point crossover with $p_{\tt cross} = 0.4$ and a mutation rate of $p_{\tt mutate} = 0.1$. The evolutionary algorithm used is the simple one described in Chapter 7 of Ref.~\cite{back2018evolutionary}.
\item {\bf $\mu + \lambda$}: This variant has the same settings as the standard one above, except that the previous generation of $\mu$ parents competes with the offspring for a place in the next generation. This is expected to help stabilize the evolution by protecting against offspring models with low fitness scores.
\item {\bf Elitist $\mu + \lambda$}: This variant has the same settings as the $(\mu + \lambda)$ algorithm above, but the best 2 solutions in a population are guaranteed to survive, 
which helps to stabilize the evolution by always retaining the fittest solutions in the population.
\item {\bf Erratic}: We also consider all three versions mentioned above, but now setting $p_{\tt cross} = 0.55$ and $p_{\tt mutate} = 0.25$. This allows the population to more aggressively move around the hyperparameter space.
\end{itemize}

\subsection{Hyperparameter intervals} \label{sec:hp_intervals}

Given a seed network architecture, the GA optimizer will explore the hyperparameter space around this seed value. Each hyperparameter's value will be restricted to a valid interval, and the tensor product of these intervals defines the optimization problem's domain. 

Let ${\bf \lambda}_{\mathrm seed}$ be the seed hyperparameter for the template network, then the hyperparameter's interval is given by
\begin{align} \label{eq:mapped_interval}
I_m = \left[ {\bf \lambda}_{\mathrm seed} (1-S_m), {\bf \lambda}_{\mathrm seed} (1+S_m)\right] \,
\end{align}
where $S_m$ is a parameter used to create an interval surrounding ${\bf \lambda}_{\mathrm seed}$. We typically set its value to $0.65$. For certain hyperparameters, we modify the lower and upper bounds of $I_m$ to comply with valid ranges (``hard limits'') as well as performing other 
necessary adjustments. For example, the ADAM optimizer's moment decay values must lie between $0$ and $1$. Similarly, for discrete variables, we would move the upper and lower limits to the nearest positive integer.

The hyperparameter's domain, $I_m$, determines both the Gaussian mutation strength and the search space of the initial population. Note that the initial population is selected from a uniform distribution on the interval $I_m$, which allows the candidate solutions to initially explore a large portion of the search space.

\subsection{Libraries and computational hardware}
\label{sec:software_hardware}

Our hyperparameter optimization algorithm (cf.~Sec.~\ref{sec:GA}) is implemented using the Distributed Evolutionary Algorithms in Python (DEAP) framework~\cite{DEAP_JMLR2012}. This framework provides for customizable classes which control the mutation, cross-over, and selection rules applied to the population of candidate models. One of the advantages of genetic algorithms is that they are easily parallelized, and we use the Mpi4py library to distribute the population over available GPU-enabled compute nodes.

We setup the model's architecture using Keras' API to the Tensorflow library. Tensorflow allows for training on GPU devices, which we make extensive use of here. Our GA simulations have typically been performed on our local cluster, CARNiE, which has 20 nodes with NVIDIA Tesla V100 GPUs. As GA optimization requires significant computational resources, having access to a GPU cluster proved to be crucial for our studies.

\section{Numerical experiments: Training and optimizing the network}
\label{sec:numerical_experiments}

Our numerical experiments will focus on exploring the properties of the genetic algorithm. We consider its performance under different scenarios and between GA variants. The goal here is not to compare with traditional matched filtering searches but rather explore the viability of hyperparameter optimization using a genetic algorithm.

\subsection{Discovering networks from scratch}
\label{subsec:lowfit}

\begin{figure*}[t!]
\centering
\begin{subfigure}[t]{0.5\textwidth}
\centering
\includegraphics[width=0.9\textwidth]{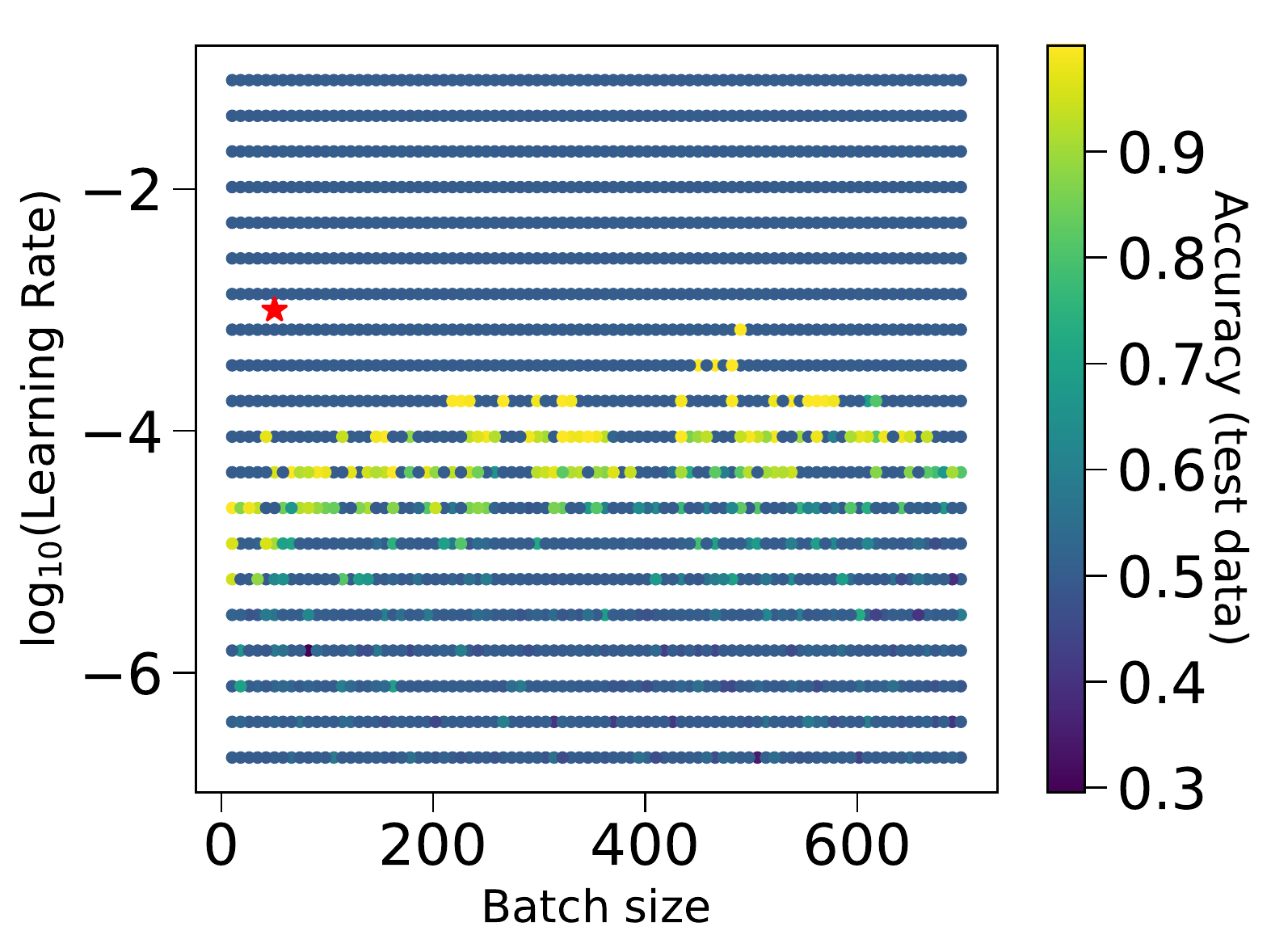}
\caption{2D grid search}
\label{fig:optimizer_2D_grid_LR_BS}
\end{subfigure}%
~ 
  \begin{subfigure}[t]{0.5\textwidth}
\centering
\includegraphics[width=0.89\textwidth]{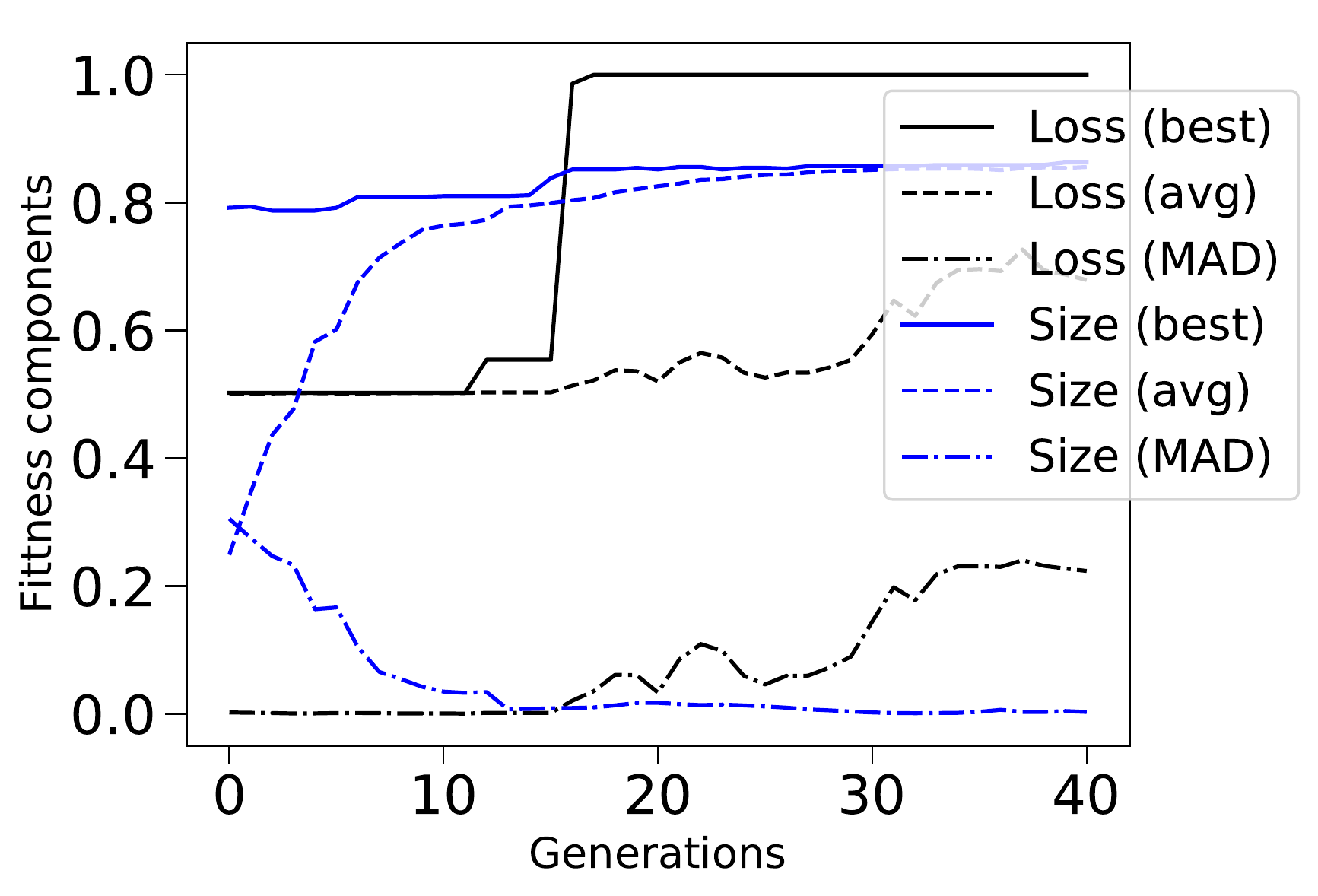}
\caption{Fitness vs generation}
\label{fig:logbook_statistics_optimizer_2D_grid_GA}
\end{subfigure}
\centering
\begin{subfigure}[t]{0.5\textwidth}
\centering
\includegraphics[width=0.9\textwidth]{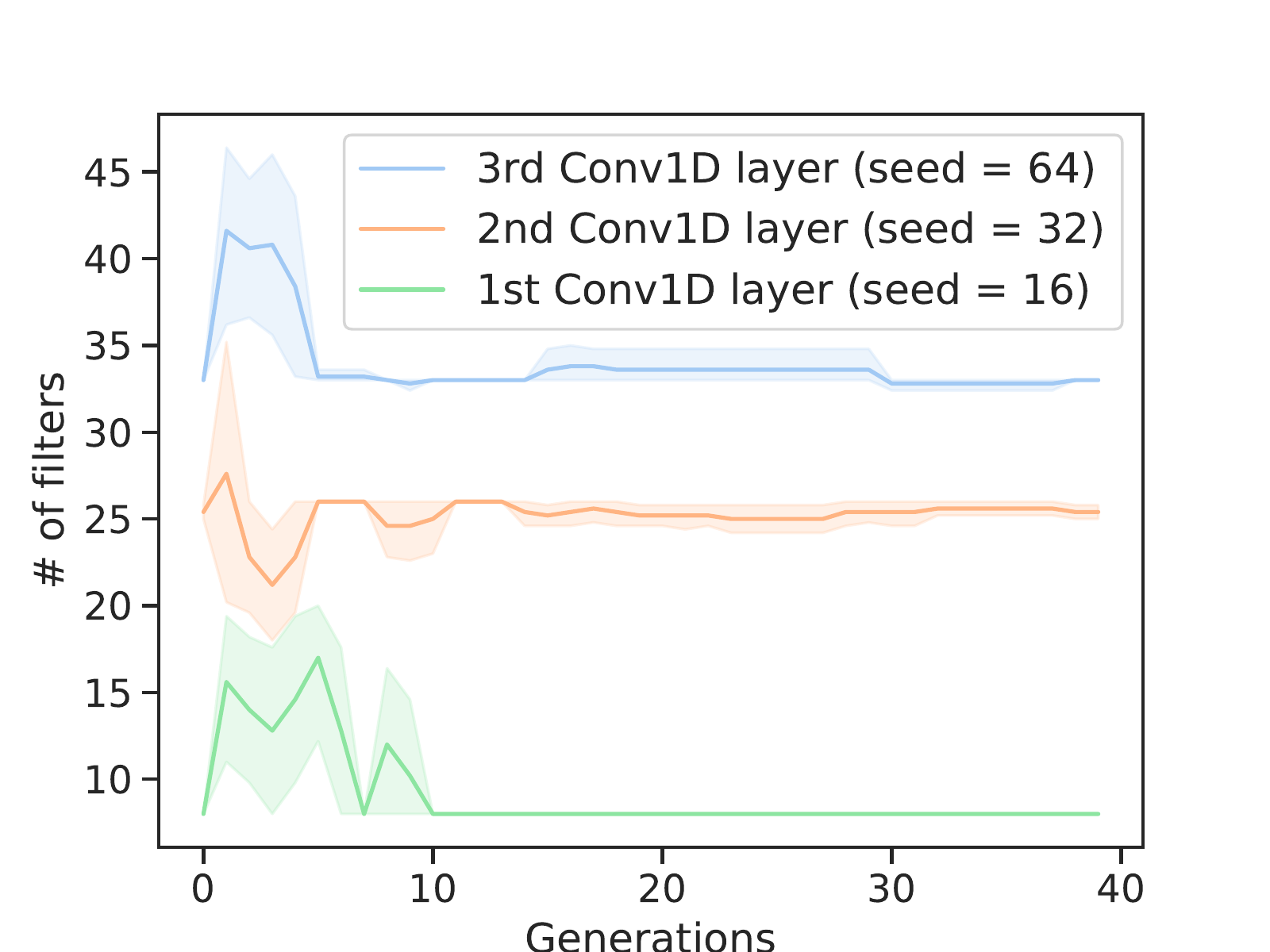}
\caption{Evolution of Conv1D filter parameters}
\label{fig:Conv1D_filter_lineplot}
\end{subfigure}%
~ 
  \begin{subfigure}[t]{0.5\textwidth}
\centering
\includegraphics[width=0.89\textwidth]{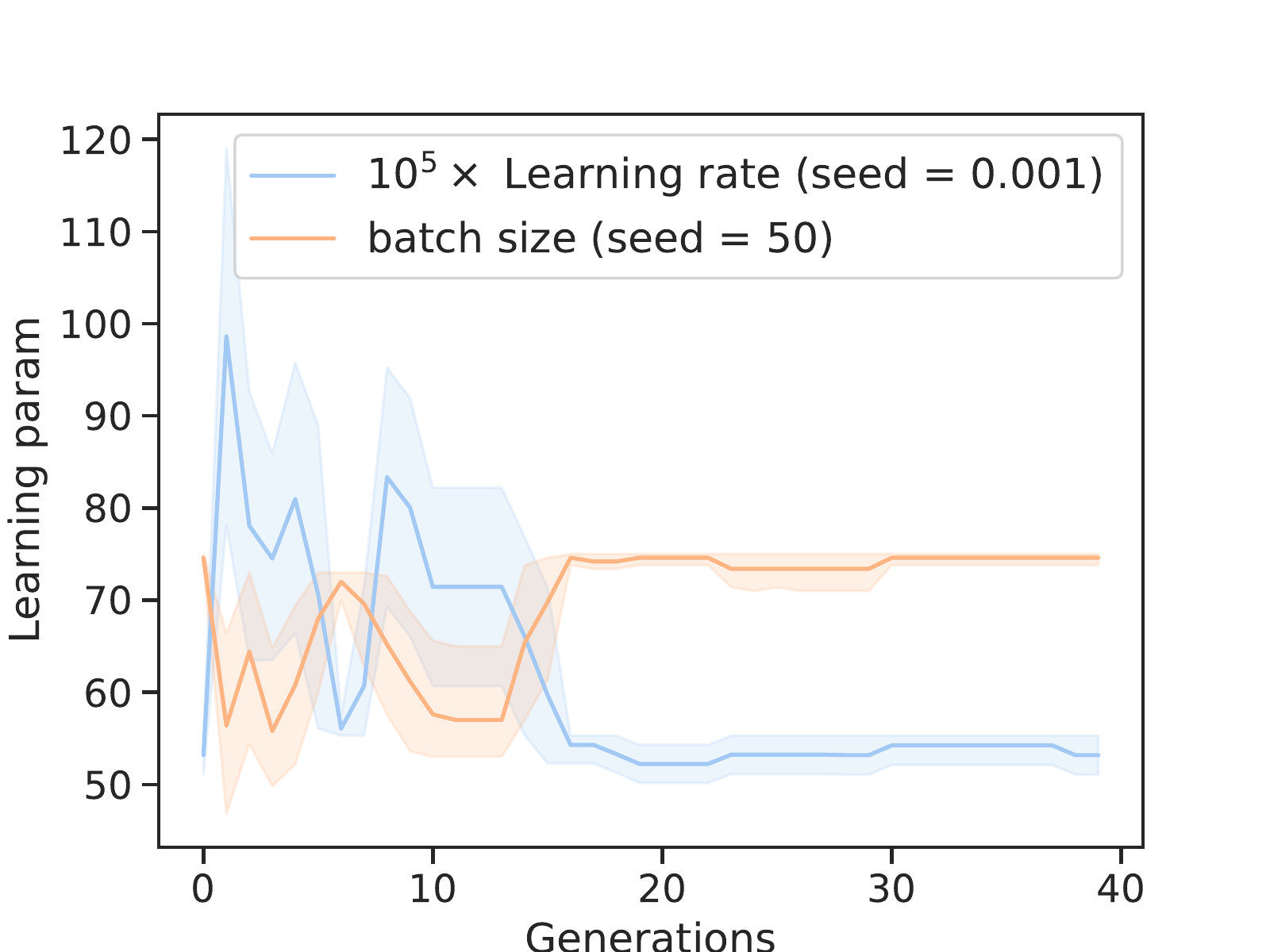}
\caption{Evolution of ADAM learning parameters}
\label{fig:Setting_ADAM_lineplot}
\end{subfigure}
\caption{These figures summarize the experiment described in Sec.~\ref{subsec:lowfit}, where we challenge the genetic algorithm by seeding the learning training in a region of low fitness (upper left panel, red asterisk). The upper right panels show that the genetic algorithm finds a region of high fitness, and the resulting network is both very accurate and much more compact than the seed network (in our definition of network complexity, a network with a complexity factor of 1 would have no learned parameters). The bottom panels show the evolution of Conv1D filters (left) and ADAM learning parameters (right) averaged over the top 5 networks. The network's seed values are shown in the legend.}
\label{fig:simple_optimization_example}
\end{figure*}

In our first example, we consider starting hyperparameter values
for which the network cannot learn at all. Most of these 
values are depicted in Fig.~\ref{Fig:classifier_network}, and 
we also select $D^i_{\tt drop} = 0.2$, $N_{\tt batch} = 50$, and 
$\epsilon_{\tt LR} =0.001$ as the seed values. As is well known, 
the ADAM optimizer can fail when the learning rate is either too 
high or too low. We have purposefully specified a large value
to show how the GA can overcome poor starting values. 

Our classification problem is defined by
$M \in [40, 60]$, 
$q \in [1, 3]$, 
$T = 1$ s,
$f_s = 2048$Hz,
Gaussian noise, 
$SNR \sim 15$, and $f_{\tt signal} = 0.5$.
Our training data is comprised of a 
few thousand examples with 20\% held out for validation. We restrict to 
high SNRs to facilitate a comparison to a dense grid-based search
on unreasonably large grids to challenge the evolutionary algorithm.

To get a better sense of the search subspace, we first perform a brute-force, grid-based search for optimal parameter values by fixing all
of the hyperparameter values except for the batch size and learning rate. A total of $1740$ unique training runs are performed on the grid depicted in Fig.~\ref{fig:optimizer_2D_grid_LR_BS}, and for each run, we stop the training sequence if the network fails to improve after 10 epochs. As the network's learned parameters are randomly initialized, sometimes the ADAM optimizer will fail simply due to unlucky initial values. And so we also retry training a failed network with new initialized values up to 4 times. Fig.~\ref{fig:optimizer_2D_grid_LR_BS} also shows how the resulting network's accuracy varies with these two parameters, where the accuracy is computed as the mean of the diagonal entries of the confusion matrix as $1 - \alpha /2 -  \beta /2$. As expected, there is a region of equally valid solutions where the classifier obtains perfect accuracy along with large regions where the network exhibits poor accuracy. Recall that since half of our training examples contain a signal, an accuracy of $0.5$ corresponds to a random guess.

Next, we solve the optimization problem using a genetic algorithm. For this experiment, we proceed with different values for the GA operators to 
show good solutions can be found without fine-tuning the GA's parameters. The GA uses $p_{\tt mutate} = 0.4$, $p_{\tt cross} = 0.6$, 
$p_{\tt gene} = 0.2$, a population size of 60, tournament size of 2, and each subsequent population retains the top two best solutions from the previous generation (the elitism selection rule). We slightly modify our weighted fitness function to weight loss by 0.95 and network complexity by 0.05.

We continue to use the seed values mentioned above while now allowing the hyperparameter values to vary over the full $17$-dimensional space. Our  genetic algorithm seeks to find accurate and compact networks by maximizing the fitness function~\eqref{eq:GA_fitness}. Fig.~\ref{fig:logbook_statistics_optimizer_2D_grid_GA} shows the accuracy and complexity contributions to the overall fitness score of the top 5 networks in the population as a function of generation. By the 17th generation we have found a network with perfect accuracy and whose size is about 85\% more compact than the original seed network. At this point we could reasonably halt the optimization algorithm, while continuing shows that future generations continue to evolve with more and more of the population moving towards regions of higher fitness. 

Figures \ref{fig:Conv1D_filter_lineplot} and \ref{fig:Setting_ADAM_lineplot} show how the hyperparameter values evolve away from their seed value and towards fitter networks. For example, Fig.~\ref{fig:Conv1D_filter_lineplot} shows that in all three convolutional layers significantly fewer filters are needed. Fig.~\ref{fig:Setting_ADAM_lineplot} shows the GA moves the learning parameters in the direction anticipated by the brute force search show in Fig.~\ref{fig:optimizer_2D_grid_LR_BS}. 

The final network discovered by the GA is more compact than others that have been reported in the literature, which is perhaps not too surprising given the comparatively smaller size of signal space. Nevertheless, this example highlights the effectiveness of the GA at automatically discovering compact and accurate networks tuned to a specific problem.

\subsection{Comparison of genetic algorithms}
\label{subsec:GA_comparison}

\begin{figure}[h!]
\centering
\includegraphics[totalheight=5.8cm]{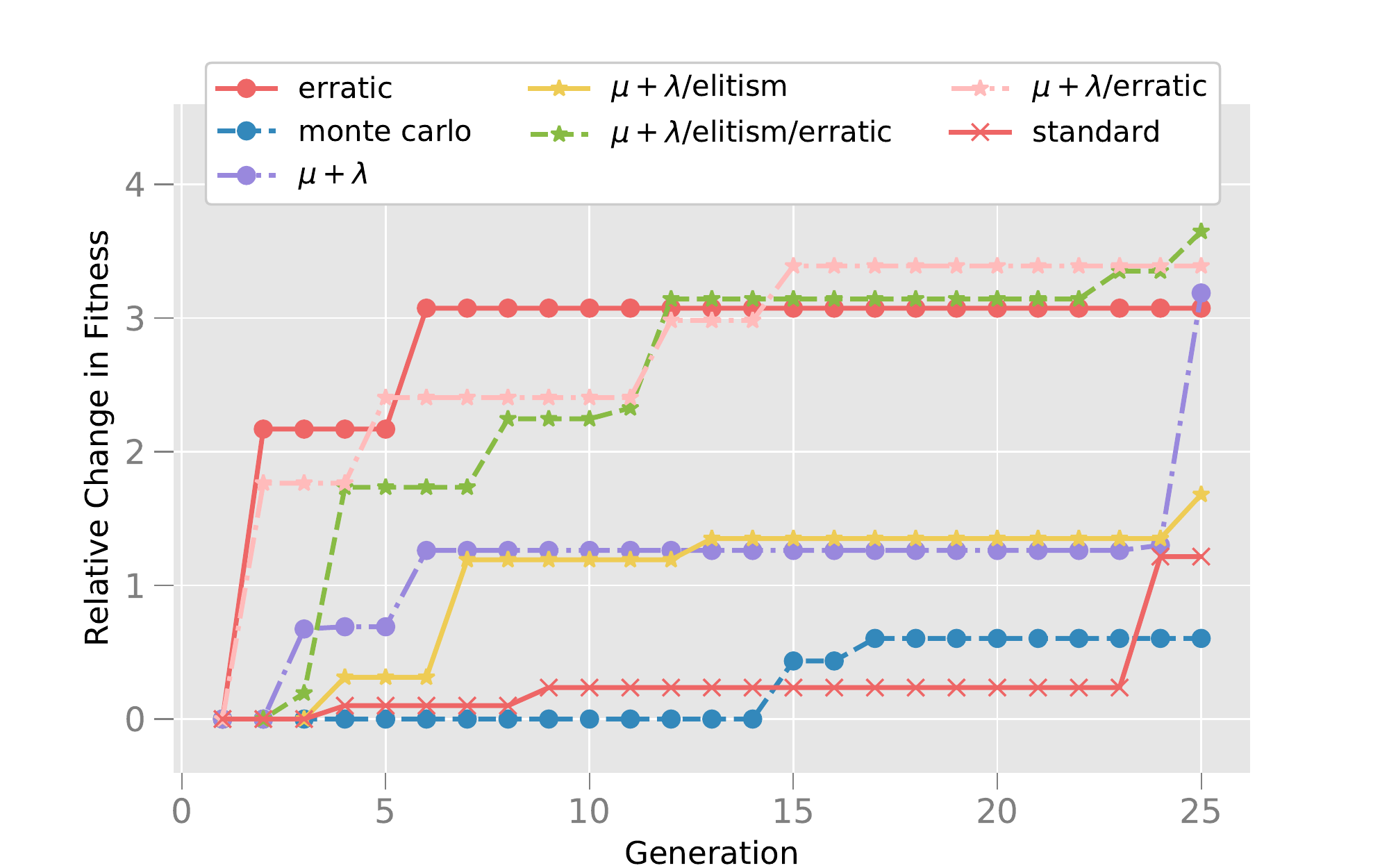}
\caption{Relative change in weighted fitness~\eqref{eq:GA_fitness} versus generation for the 6 different genetic algorithm variants considered here. The weighted fitness, which includes contributions from both the network's accuracy and complexity into the objective function, is the quantity that the GA optimizes over this 20-dimensional hyperparameter search space. We see that GA variants that explore the parameter space most aggressively (erratic versions) provide for continual refinements of the population through all 25 generations. For comparison, we also show random sampling (Monte Carlo) of the hyperparameter space, which, given the high dimensionality of the problem, results in many generations without any better performing candidates. While the improvements in all cases are modest, we note that the GA is refining the George \& Huerta architecture whose hyperparameter values have been found through an extensive, manual, and randomized trial-and-error procedure.}
\label{Fig:weighted_fit}
\end{figure}

\begin{figure}[h!]
\centering
\includegraphics[totalheight=5.8cm]{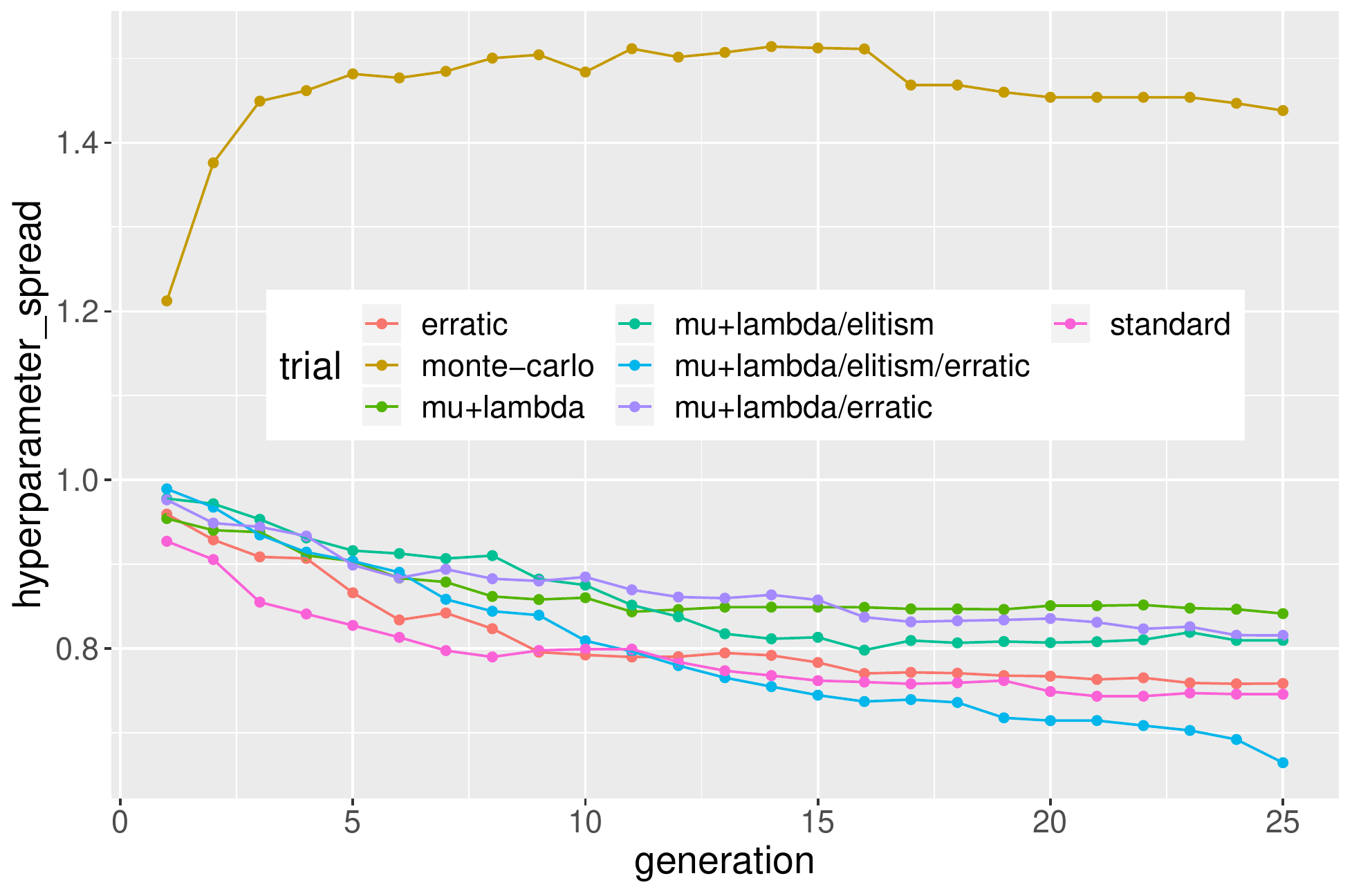} 
\caption{Standard deviation in the (normalized) hyperparameter values from the top 10 solutions. In all trials, our GA's appear to converge to a minima as shown by the decreased diversity in the population of solutions. Note that because we use 5 non-interacting populations the spread of hyperparameter values is larger than the spread in each trial (cf. Fig.~\ref{fig:final_spread_conv}). Notice that after many generations most GA populations slow their evolution while $\mu+\lambda$/elitism/erratic continues to find new, high-quality solutions. The top 10 solutions found through random sampling (Monte Carlo) do not show any similarity, which indicates that there are a variety of decent solutions for this problem.}
 \label{fig:comp_spread_converg}
\end{figure}

In the previous example, we considered how a genetic algorithm could improve on hyperparameter values for which the seed network cannot learn at all. We now consider how genetic-algorithm optimization can improve upon an already good network architecture as well as comparing GA variants. 

In this first numerical experiment, we perform hyperparameter optimization with the six different genetic algorithm variants described in Sec.~\ref{sec:GA_variants} as well as random sampling (Monte Carlo). Each GA uses the same network seed shown in Fig.~\ref{Fig:classifier_network}. 
Our key aim is to compare the convergence and fitness properties of each approach as well as understanding how these algorithms move through the hyperparameter space. In all cases, the search space is defined by the hyperparameter ranges quoted in Table~\ref{tab:hparams}. 

We use a sequence of 15 training datasets with 2000 training examples per dataset. Each training example corresponds to $T =1$ second of time series data sampled at a rate of $\Delta t = 2048$ Hz. In our noise model, we set a constant value of $S_n(f_i) = S_n$ (cf.~Sec.~\ref{sec:signal_detection_setup}) such that each dataset in the sequence contains signals with a typical target SNR computed from Eq.~\eqref{eqn:maxamp_stat}. Our sequence of datasets is constructed such that the classification problem is gradually more challenging, which is achieved by gradually lowering the SNR from 1828.272 to 2.711 at a decelerating rate (cf.~Sec.~\ref{sec:training_set_hparams}). The last 4 datasets in this sequence have signals with average SNR values of 14.7961, 9.9239, 6.9361, and 2.711. From these four datasets we hold out 20\% of the training examples to compute the accuracy fitness score~\eqref{eq:GA_fitness}.

Gravitational-wave signals are simulated using a non-spinning numerical relativity surrogate model~\cite{Blackman:2015pia}. We simulate systems by sampling a uniform distribution with the total mass from 22.5 to 67.5 solar masses, the mass ratio from 1 to 5, and the distance from 0.5 to 1.5 megaparsecs. The binary system is oriented such that $\iota = \phi_c = \pi / 3$, and we choose values of right ascension (ra), declination (dec), and polarization ($\psi$) such that the antenna patterns satisfy $F_+ = 1$ and $F_\times = 0$. Half of the training examples are pure noise and half contain a signal. Signals are added to the noise with random time shifts such that the signal's peak amplitude occurs at different times. 

In general, optimization algorithms will tend to converge to a local minima. In the case of the GA, there are a variety of strategies to overcome this problem including simulated annealing~\cite{thangiah1994hybrid,gandomkar2005combination}, dual population algorithms~\cite{park2010dual}, and others. We instead follow a more brute-force approach by rerunning the algorithm 5 times, with each run using the same seed network but 20 with distinct candidates in the initial population. All 5 populations are advanced forwards over 25 generations. We then aggregate these 5 distinct (non-interacting) sub-populations into a single population of size 100. As there are 6 distinct GA variants plus Monte Carlo tested here, we have trained a total of 17,500 CNNs to compile the results of this subsection. 

Figures~\ref{fig:comp_spread_converg} and \ref{Fig:weighted_fit} show the evolution of these $100$ candidates over all 25 generations. This provides us with an overview of how each GA algorithm is performing at the expense of a more detailed view of each of the 5 sub-populations. Note that since each sub-population may be converging towards a local minima, large hyperparameter spreads should be interpreted as sub-populations converging to different parts of the parameter space. Later on in Sec.~\ref{sec:FinalGA} we explore a more detailed view of one particular genetic algorithm.

We first consider how well each GA performs its primary task, which is to optimize the weighted fitness objective function given in Eq.~\eqref{eq:GA_fitness}. Fig.~\ref{Fig:weighted_fit} compares 6 GA variants for this problem by plotting the change in the highest achieved fitness for any individual network (i.e. the current best solution) versus generation. Due to the random starting values of the initial population there is already some spread amongst GA variants at the first generation. To account for this we monitor the relative percentage change, $100 \times | \max_i S_i^j - S_{\tt seed} | /  S_{\tt seed}$, from the initial weighted fitness value, $ S_{\tt seed} = \max_i S_i^0$. To assess the algorithm's performance, we consider how quickly the algorithms can achieve higher fitness scores. While all of the variants provide good performance on this challenging (low SNR) case, it is clear that GAs that more aggressively explore the parameter space (``erratic" versions) offer better performance. In particular, we see that erratic $\mu+\lambda$ with elitism continues to find refined hyperparameter values throughout the simulation. Non-erratic versions, for example the standard and $\mu+\lambda$ variants, are less effective at exploring the space and are characterized by no improvement for multiple generations. Monte Carlo sampling approach also fails to find better candidates for most of the simulation. 

Next we turn our attention to how the hyperparameter values evolve. If the optimization problem has neither local minima nor degeneracies then we would expect to see the population converge to a unique point in parameter space. Consequently, under this scenario, we would expect the average spread of hyperparameter values to converge towards zero. For the complicated problem considered here, however, we instead expect potentially many local minima and degeneracies. Additionally, as described above, we have combined results from 5 non-interacting populations each of which might converge to different local minima.  Nevertheless, it is still useful to monitor the diversity of the entire population over generations. Fig.~\ref{fig:comp_spread_converg} shows the average (over all 20 hyperparamters) standard deviation of the (normalized) hyperparameter values in top 10 best solutions. We find that all of the genetic algorithm variants show some form of convergence that tends to slow with generation. The $\mu+\lambda$/elitism/erratic GA variant shows the fastest convergence of the 6 variants, which was also seen in the fitness plot Fig.~\ref{Fig:weighted_fit}. Due to Monte Carlo's global, uncoordinated sampling, there is very little similarity seen among the best solutions for this case. 

One common measure of algorithm performance is the rate at which it converges toward the solution, in this case, the maximum value of the fitness function. For genetic algorithms there are some theoretical results on convergence~\cite{sharapov2006convergence,eiben1990global,cerf1998asymptotic}, but its not clear how applicable these results are to our case. For this problem, empirical evidence from Fig.~\ref{Fig:weighted_fit} indicated that erratic $\mu+\lambda$ with elitism performs at least as good as the other GA variants, and in some cases much better. This is possibly due to the synergy between the stabilizing properties of $\mu+\lambda$ with elitism combined with the aggressively explorative search rates used in erratic variants. We did find, however, that for small populations sizes (less than 10) genetic algorithms with elitism are more susceptible to local minima, and so in this regime the stabilizing features are counterproductive. We also tested erratic $\mu+\lambda$ with elitism using a two-point crossover and mutation operators, but these more aggressive search operators had little effect and so these results were not included here.

\begin{figure*}
\centering
    \begin{subfigure}[b]{0.45\linewidth}
        \includegraphics[width=\linewidth]{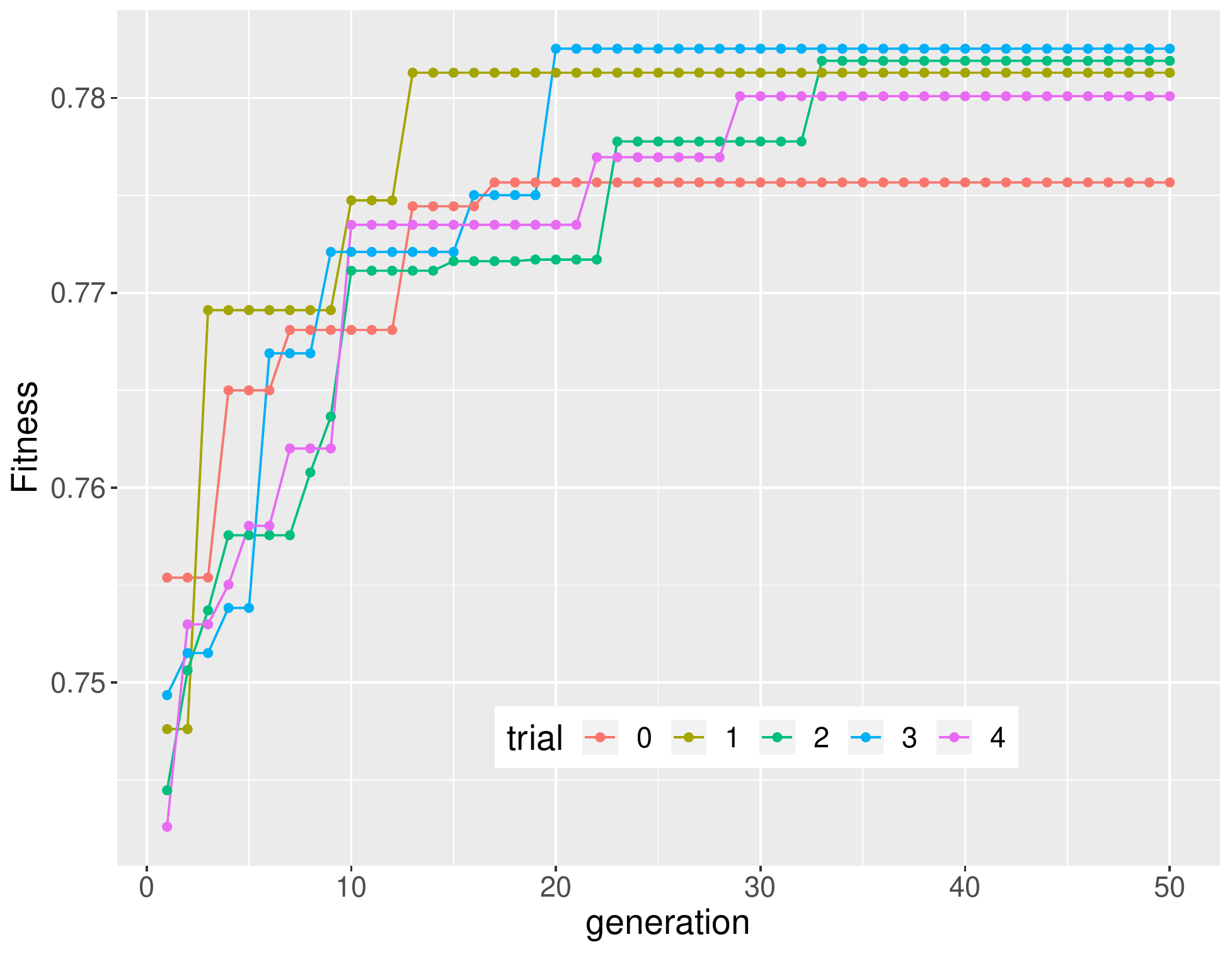}
        \caption{Weighted fitness (seed value = 0.68)}
        \label{fig:final_weighted_fit}
    \end{subfigure}
    \begin{subfigure}[b]{0.45\linewidth}
        \includegraphics[width=\linewidth]{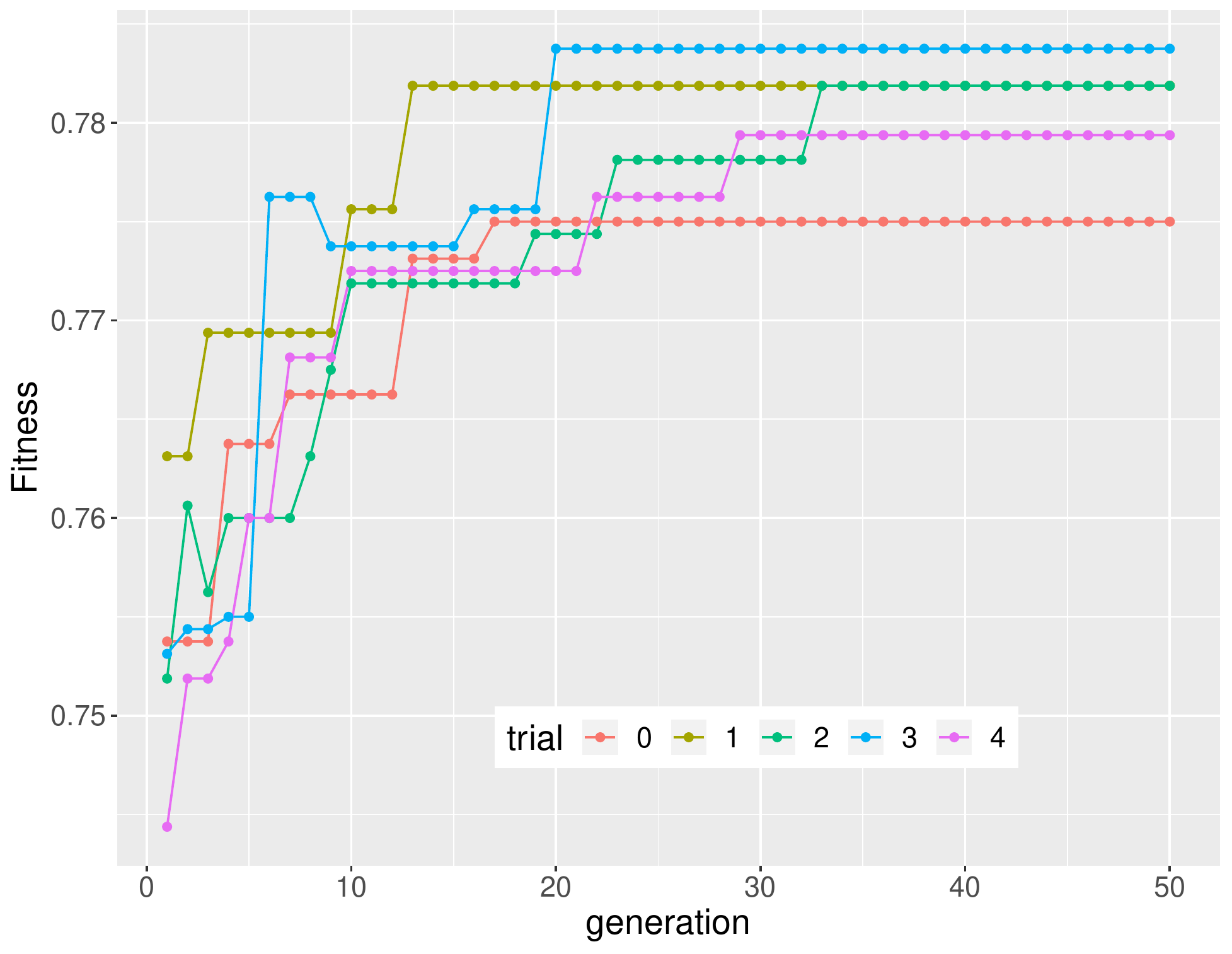}
        \caption{Accuracy fitness (seed value = 0.7)}
        \label{fig:final_loss_fit}
    \end{subfigure}
    \begin{subfigure}[b]{0.45\linewidth}
        \includegraphics[width=\linewidth]{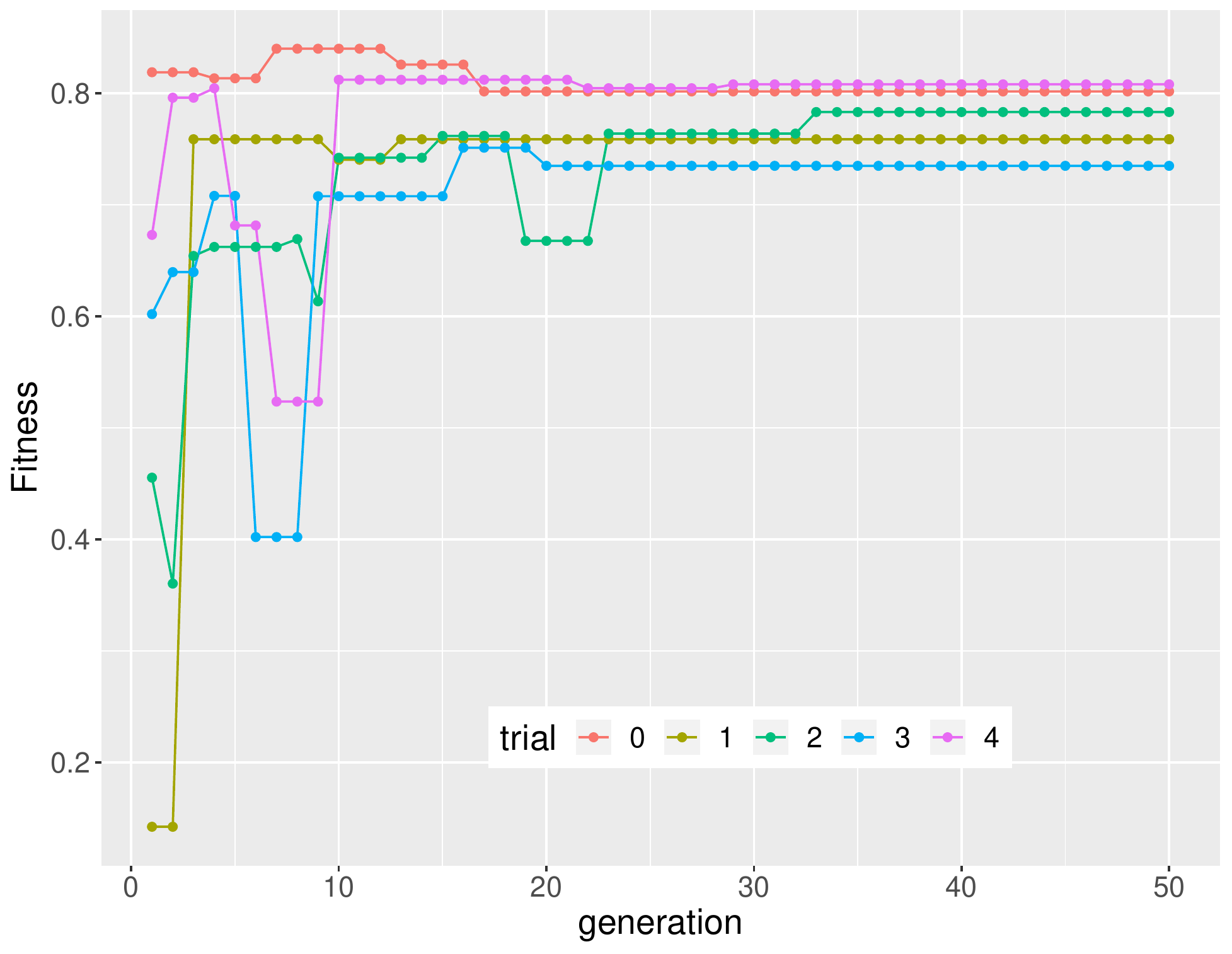}
        \caption{Complexity fitness (seed value = 0)}
        \label{fig:final_size_fit}
    \end{subfigure}
    \begin{subfigure}[b]{0.45\linewidth}
        \includegraphics[width=\linewidth]{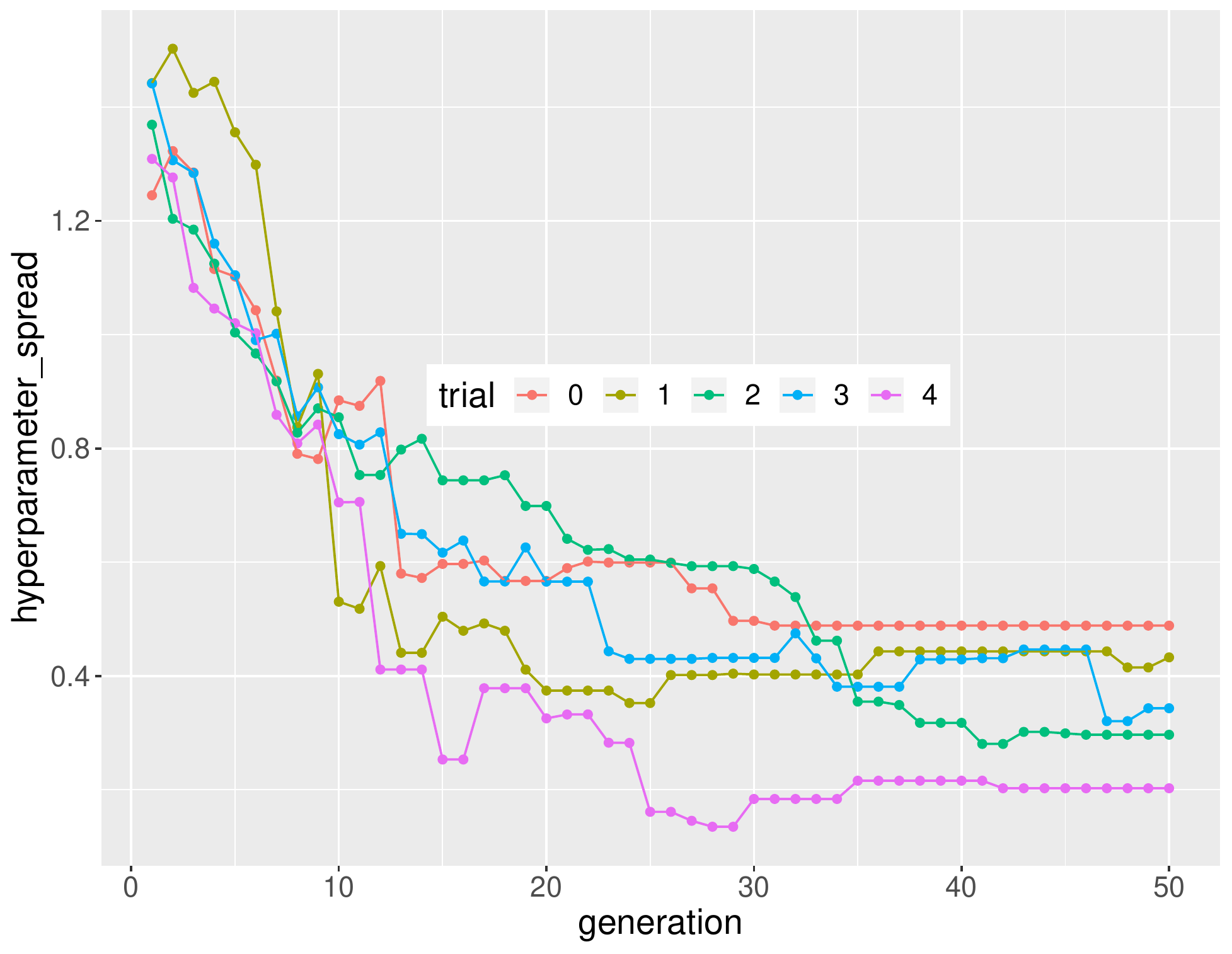}
        \caption{Hyperparamter spread vs generation}
        \label{fig:final_spread_conv}
    \end{subfigure}
    \caption{These figures summarize the experiment described in Sec.~\ref{sec:FinalGA}, where we perform hyperparameter optimization with the erratic $\mu+\lambda$ with elitism GA variant using five distinct 
    runs (labeled trials) to help guard against local minima. In all trials, the populations appear to have converged to a local minima by about 35 generations. This is most evident by monitoring the spread in the (normalized) hyperparameter values in the top 10 solutions (bottom right). 
    The genetic algorithm finds a region of high weighted fitness (upper left panel) and the resulting network is both 
    more accurate (upper right panel) and more compact (lower than panel) than the George and Huerta small classifier (our seed network), whose values are shown in the subfigure's caption. The GA made significant improvements in network complexity, which had about 80\% fewer learned parameters as compared to the seed network. We have included extremely weak signals in our validation set with SNRs as low as 2, which is why our accuracy fitness obtains a maximum of $\approx 79\%$; Sec.~\ref{sec:network_comparisons} explores network properties as the SNR is varied.
}
\label{fig:simple_optimization_example}
\end{figure*}

It is worth noting that certain algorithms are more computationally challenging to run. For example, at each generation the Monte-Carlo algorithm, the most costly of the algorithms we considered, selects an entirely new set of individuals all of which need to be retrained. By comparison, the fittest individuals in the GA population are carried over to the next iteration and do not need to be trained. However, this difference is more significant for cpu- or gpu-time than walltime since training a population of networks is embarrassingly parallel, unlike generations that proceed sequentially.

\subsection{Network refinement using erratic $\mu+\lambda$ with elitism}
\label{sec:FinalGA}

\begin{figure}[h!]
\centering
        \includegraphics[totalheight=7.0cm]{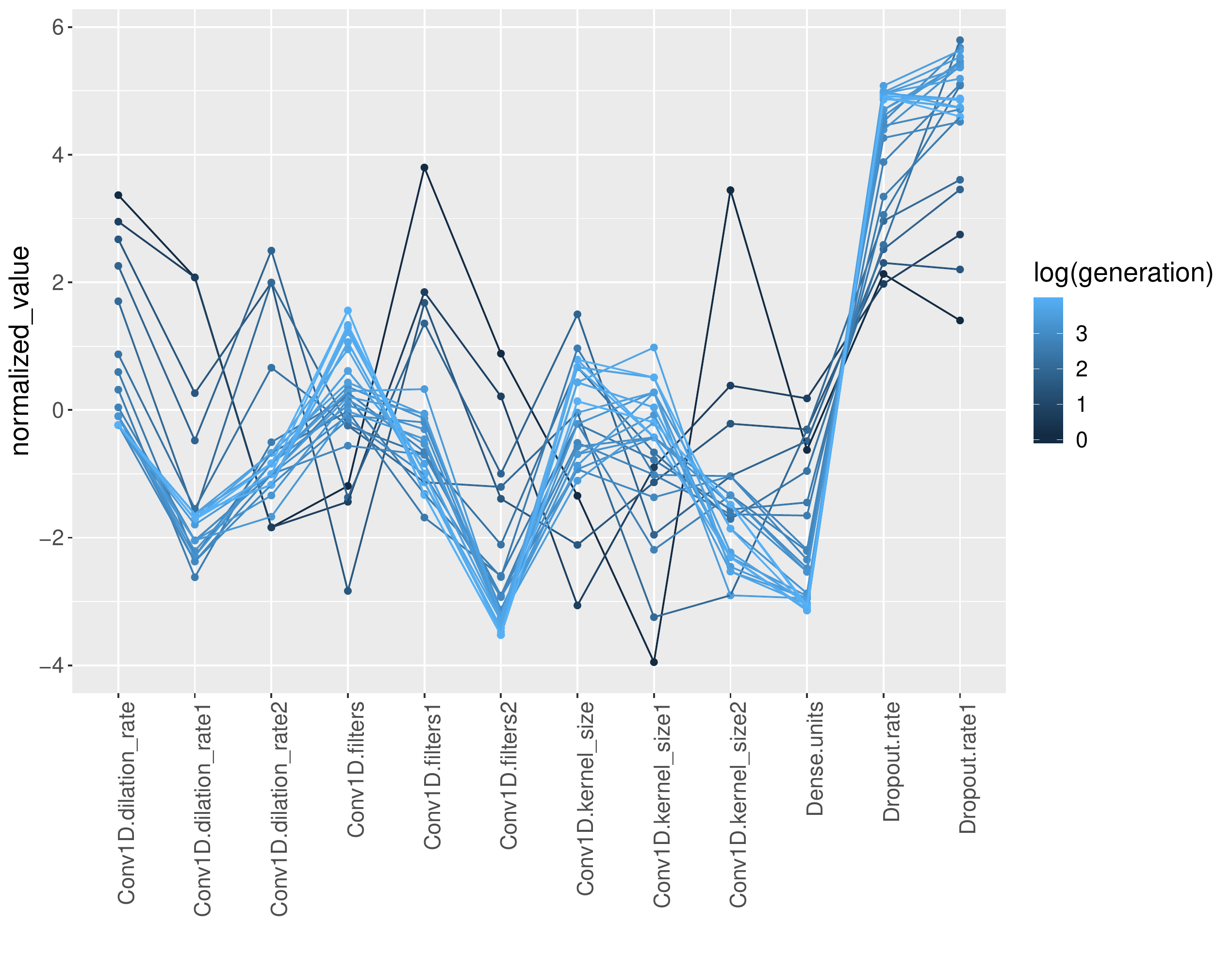}
        \caption{Parallel coordinate plot of hyperparameter evolution over 30 generations of architecture optimization using the GA-variant 
        described in Sec.~\ref{sec:FinalGA}. The generations are shown logarithmically to better distinguish earlier generations from one another. Each generation is represented as a color-coded, connected line showing the 
        best performing architectures averaged over the 5 sub-populations. 
        Hyperparameter values are normalized such that a value of 0 corresponds to value of the George \& Huerta seed architecture, thus allowing us to compare GA's solution to the seed network as well as the evolution the population took to arrive at the optimized architecture.
}
\label{fig:final_hparam_value_conv}
\end{figure}

In Sec.~\ref{subsec:GA_comparison} we found that the GA variant erratic $\mu+\lambda$ with elitism was a top performer for this problem. Here we explore this GA variant in a bit more detail while scaling up the search to use a population size of 50, 50 generations, and 5 elites (the elites have been scaled proportionally to the population). We continue to use the same training and testing data as in Sec.~\ref{subsec:GA_comparison}. Also as before, to avoid local minima we use 5 non-interacting populations (called trials below) with different seeds, however they are now displayed separately rather than aggregated.

In Fig.~\ref{fig:simple_optimization_example} shows the evolution of the best solution for each trial. We see that by about 35 generations each trial has converged to a solution that has improved upon the seed network, whose fitness values are shown in each subfigure. It should be noted that although our GA will guarantee improvements in weighted fitness (cf. Fig.~\ref{fig:final_spread_conv}), accuracy and complexity fitness have no such guarantee since the algorithm is optimizing the weighted fitness. Figure~\ref{fig:simple_optimization_example} shows that both fitness measures increase. 

From Fig.~\ref{fig:final_weighted_fit} we see that trial 3 found the network with the best overall accuracy, whose accuracy fitness improved from .71 (accuracy of the seed network) to .79, an 11\% increase. The accuracy baseline of the seed network was computed by taking the best score after retraining 10 times from scratch to guard against unlucky weight initializations. By comparison, due to computational cost considerations, each member of the GA's population was only evaluated once with a randomized seed. Due to a large number of generations and population size the GA effectively explores many possible seed choices to make the impact of an unlucky seed unimportant. 

Fig.~\ref{fig:final_loss_fit} shows that this network also has 78\% fewer learned parameters as compared to the seed network. Note that in our definition of complexity, a value of 0 means the network has as many learned parameters as the seed network while a value of 1 is a trivial network with no trainable parameters. We remind the readers that our seed network architecture was taken to be the best network with 3 CNN layers from Ref.~\cite{George2018FirstPaper}, and so we see here the ability of the GA to improve upon already good networks, which will be important for maximizing the efficacy of machine-learning based gravitational wave searches.  

Finally, in Fig.~\ref{fig:final_hparam_value_conv} we show the evolution of (normalized) hyperparameter values across generations. This provides some insight into the influence of a given hyperparameter for this problem. For example, the 2nd and 3rd convolutional layers' dilation rates benefited from being smaller. The dense units also moved to notably smaller sizes while higher dropout rates were preferred. We also observe a decrease in Conv1D kernel sizes from layers 1 to 3, perhaps since the maxout layers reduce the activation areas between each convolutional layer. There also appears to be degeneracy among the first two filter values, where the population wanders between many plausible values even at later generations, however, the last filter value tends to become smaller. 

\begin{table}
\caption{Genetic algorithm optimized architecture for the gravitational-wave classification problem defined in Sec.~\ref{sec:FinalGA}. For comparison, we also show the values used for the seed network, which is essentially the classifier discovered by George and Huerta~\cite{George2018FirstPaper} and shown in Fig.~\ref{Fig:classifier_network}. Interestingly, the GA was able to find a significantly more compact network that simultaneously achieves better accuracy. It is surprising how few filters and neurons the network needs; the largest number of filters and neurons per layer being only 30 and 38 respectively. It is also noticeable that the kernels in the early layers need to be very wide and relatively dense, while later kernels shrink and become sparse (as indicated by dilation) rather quickly. This is somewhat contrary to the conventional wisdom of CNN architectures and underscores the potential benefits of automated hyperparameter tuning.
}
   \label{tab:final_results}
   \centering
\begin{tabular}{c | c c}
Quantity & Seed & GA\\
\hline
Weighted Fitness             & $0.6830484375$ & $0.7825329$  \\
Accuracy Fitness                 & $0.7005625$ & $0.7837500$ \\ 
Size Fitness                 & $0$ & $0.7350671$ \\ 
CNN-1 Filters        & $16$ & $20$ \\ 
CNN-1 Size           & $16$ & $14$ \\ 
CNN-1 Dilation  & $1$  & $1$ \\ 
CNN-2 Filters        & $32$ & $14$ \\ 
CNN-2 Size           & $8$  & $6$ \\ 
CNN-2 Dilation & $4$    & $3$ \\
CNN-3 Filters       & $64$   & $30$ \\ 
CNN-3 Size          & $8$    & $4$ \\ 
CNN-3 Dilation & $4$    & $2$ \\ 
Dropout-1 Rate           & $0.1$  & $0.1948836$ \\ 
Dense Units                 & $64$  & $38$ \\ 
Dropout-2 Rate           & $0.1$  & $0.1058403$ \\
GW comparisons           & \multicolumn{2}{c}{ Sec.~\ref{sec:network_comparisons}} \\
\end{tabular}
\end{table}

\subsection{Comparing seed and optimized architectures}
\label{sec:network_comparisons}

\begin{figure*}[t!]
\centering
\begin{subfigure}[t]{0.5\textwidth}
\centering
\includegraphics[width=0.9\textwidth]{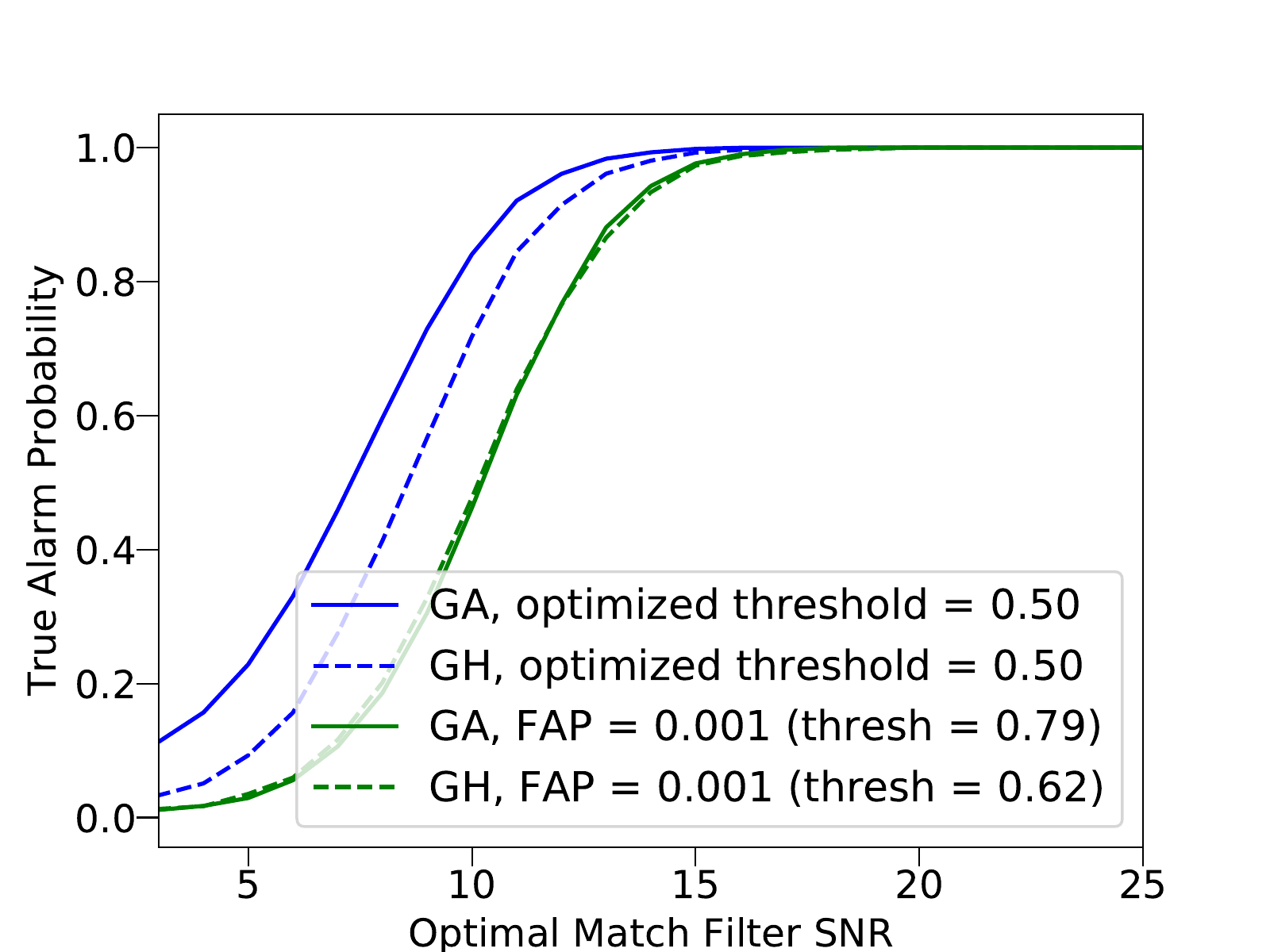}
\caption{Efficiency Curves}
\label{fig:efficiency_curves_GA_vs_GH}
\end{subfigure}%
~ 
  \begin{subfigure}[t]{0.5\textwidth}
\centering
\includegraphics[width=0.89\textwidth]{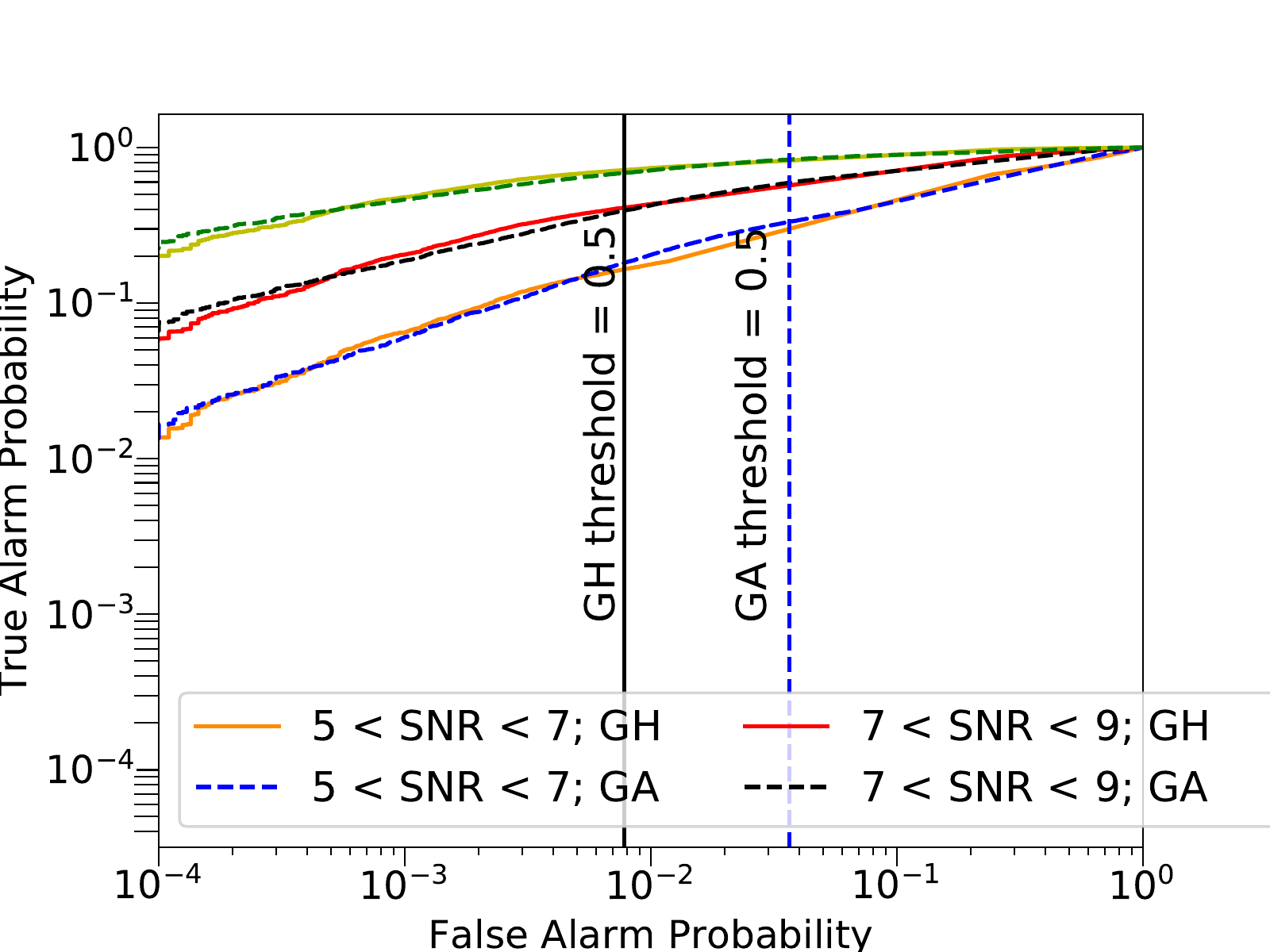}
\caption{ROC Curves}
\label{fig:roc_curves_GA_vs_GH}
\end{subfigure}
\centering
\begin{subfigure}[t]{0.5\textwidth}
\centering
\includegraphics[width=0.9\textwidth]{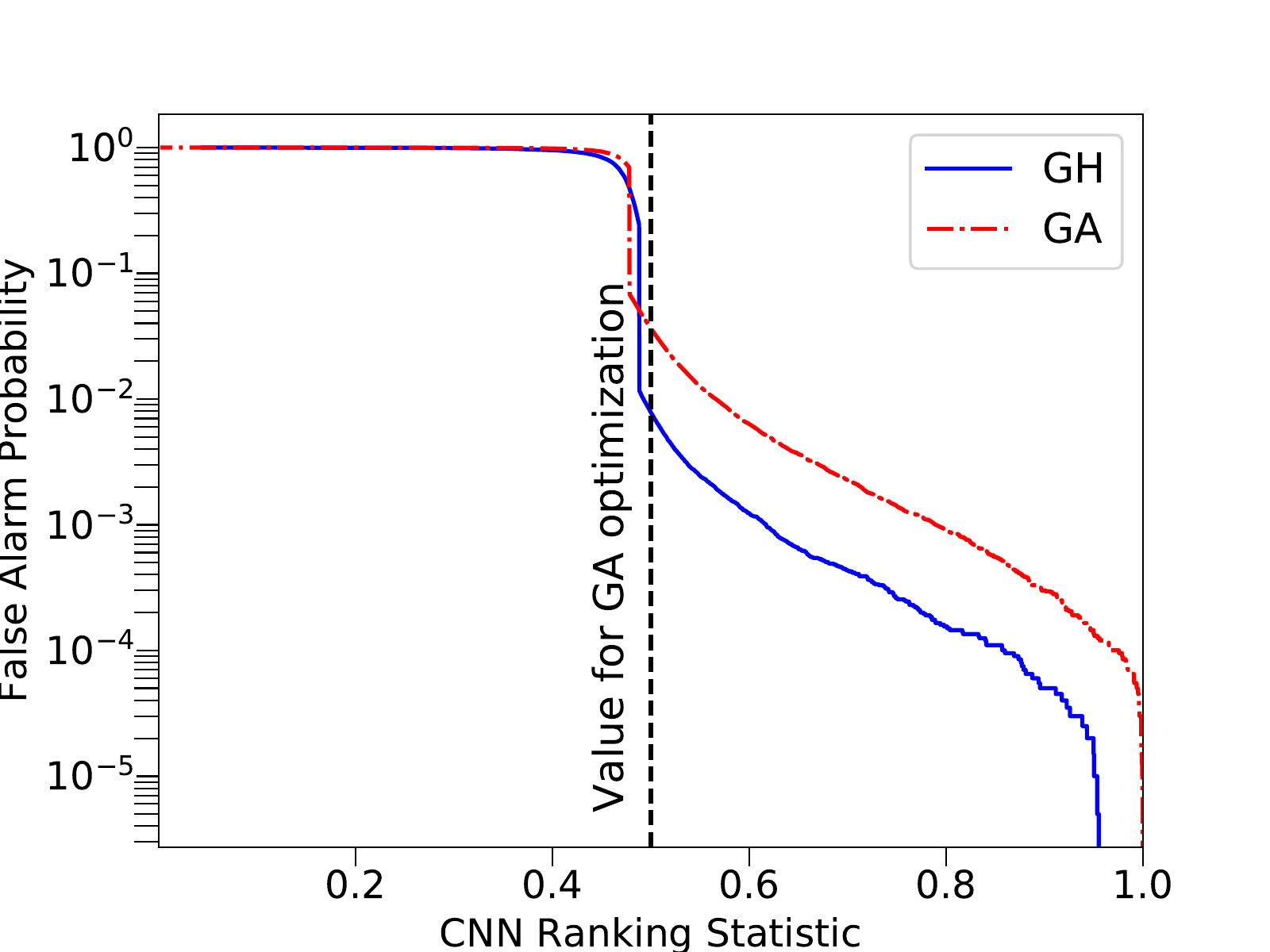}
\caption{FAP vs thresholds}
\label{fig:FAP_vs_thresholds}
\end{subfigure}%
~ 
  \begin{subfigure}[t]{0.5\textwidth}
\centering
\includegraphics[width=0.89\textwidth]{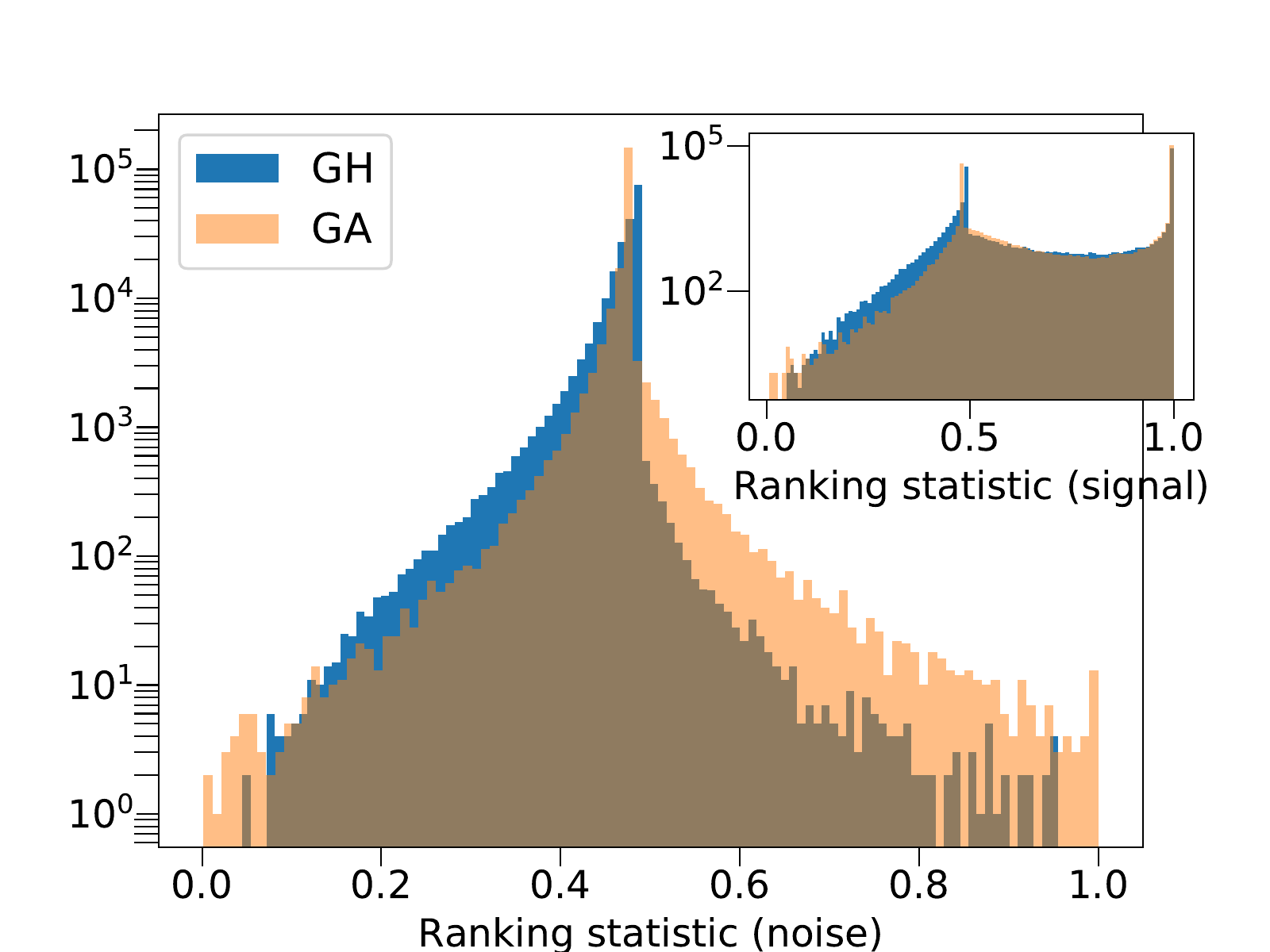}
\caption{Distribution of ranking statistics}
\label{fig:statistic_distributions_GA_vs_GH}
\end{subfigure}
\caption{Comparison of our seed network to the GA-optimized architecture as described in Sec.~\ref{sec:network_comparisons}.
In Sec.~\ref{sec:FinalGA} we showed that the optimized architecture has 79\% fewer trainable parameters while its classification accuracy is 11\% higher than the seed network when using a ranking statistic threshold of $P^*_{\tt signal}=0.5$. As seen in the upper left panel, when using this threshold value, the optimized network's true alarm probability is higher across a range of signal SNR values. However, at fixed false alarm probabilities (FAP) the optimized network does not show any clear advantage beyond being significantly more compact. This is perhaps not surprising since the fitness score~\eqref{eq:GA_fitness} uses a fixed threshold of $P^*_{\tt signal}=0.5$ instead of a fixed the FAP. Indeed, from the bottom two figures we see that the ranking statistic of the optimized network has a dramatically different distribution before and 
after the threshold value of 0.5. Exploring alternative GA fitness scores to address this issue will be considered in future work.
}
\label{fig:GA_GH_comparisons}
\end{figure*}

In Sec.~\ref{sec:FinalGA} a genetic algorithm was used to improve George and Huerta's proposed architecture (our seed network) for the 
classification problem described in Sec.~\ref{subsec:GA_comparison}. We now consider a more detailed comparison between the George and Huerta (GH)
and GA-optimized architectures by considering diagnostics discussed in Sec.~\ref{sec:diagnostics}. Our testing data is comprised of  $400,000$ examples with half containing signals with SNRs between 2 and 30. 

Recall that the output of a classier is a number, the ranking statistic, that assigns a measure of confidence that the dataset contains a signal. For traditional matched filtering this number, $\rho$, is the signal-to-noise ratio. For the CNN classifier the network outputs a number, $0 \leq P_{\tt signal} \leq 1$, which we would like to interpret as the significance of a signal: when $P_{\tt signal} = (0) 1$ the network is absolutely certain is (no) signal. The inset of Fig.~\ref{fig:statistic_distributions_GA_vs_GH} shows the distribution of $P_{\tt signal}$ over $200,000$ datasets that contain a signal. Two peaks are evident. The largest one, located at $P_{\tt signal} = 1$, corresponds to high SNR events. A secondary peak, comprised of moderate SNR events, lies just above $0.5$. The distribution of $P_{\tt signal}$ over $200,000$ noise-only datasets also shows a large peak just below 0.5. 

Following Gabbard et al.~\cite{gabbard2018matching} our next comparison between the GH and GA-optimized networks considers the true alarm probability versus the optimized matched-filter SNR of the signal. Fig.\ref{fig:efficiency_curves_GA_vs_GH} shows that the GA-optimized network (solid blue line) outperforms the seed network (dash blue like) at the threshold value of $P^*_{\tt signal} = 0.5$ used in the computation
of the accuracy when computing the GA's fitness score. However, from Figs.~\ref{fig:roc_curves_GA_vs_GH} and \ref{fig:FAP_vs_thresholds} we see that at this threshold value the networks have different false alarm probabilities. At a fixed FAP of $10^{-3}$ (green) both networks show comparable performance. We believe this is a consequence of using a threshold of $P^*_{\tt signal} = 0.5$ in the computation of the fitness score,
which, as we have empirically shown, does not control the FAP. In future work we hope to explore different loss functions or GA fitness scores
to directly control and optimize for target FAPs.

Finally, in Fig.~\ref{fig:roc_curves_GA_vs_GH} we compute ROC curves for three representative optimal matched-filter SNR values. We see that the GA outperforms the seed network for weaker signals at FAPs corresponding to thresholds near 0.5, while at higher SNR values and/or different FAPs
neither network has a clear advantage. We note that despite both networks having comparable effectiveness in some of our tests, the optimized architecture is able to achieve these results with 79\% fewer learned parameters. 

\section{Discussion \& Conclusion}

We have presented a novel method for optimizing the hyperparameter values of a deep convolutional neural network classifier
based on genetic algorithms. We have applied our method to optimize deep filtering~\cite{George2018FirstPaper} networks, a special kind of convolutional neural network classifier designed to rapidly identify the presence of weak signals in noisy, time-series data streams. For example, deep filtering has been used to search for gravitational-wave signals~\cite{George2018FirstPaper,shen2019deep,hezaveh2017fast,levasseur2017uncertainties,ciuca2019convolutional,gabbard2018matching,shen2017denoising,george2017glitch,george2018deep,fort2017towards,George2018FirstPaper,gebhard2019convolutional} 
as an alternative to more traditional (and computationally expensive) matched filtering. All previous attempts to optimize deep filtering hyperparameter values have relied on trial and error strategies to set the hyperparameter values.

The principal contribution of our work is to assess the benefits of genetic algorithms for hyperparameter optimization. Our work also constitutes the first attempt to automate the hyperparameter optimization procedure for such networks. We have specifically focused on (i) assessing the effectiveness of different genetic algorithm variants for our problem, (ii) quantifying the genetic algorithm's ability to improve upon state-of-the-art architectures, and (iii) considering the genetic algorithm's ability to discover new architectures from scratch. We also provide a detailed comparison of our fully optimized network with the network described in Ref.~\cite{George2018FirstPaper}. Our main findings include:

\begin{itemize}
  \item (i) In Sec.~\ref{subsec:GA_comparison} we compared six different GA algorithms, differing in their choice of selection, mutation, and crossover operators. While many performed comparably well, the variant erratic $\mu+\lambda$ with elitism was generally found to work the best. This is possibly due to the synergy between the stabilizing properties of $\mu+\lambda$ with elitism combined with more aggressive search operators used in our erratic variants. We also considered a GA fitness score~\eqref{eq:GA_fitness} based on both classification accuracy and network complexity, and found that the network complexity fitness term resulted in significantly more compact networks without sacrificing accuracy.
  \item (ii) Previous attempts at hyperparameter optimization relied on trial and error (Monte Carlo) searches. For the benchmark cases considered here, we find that all GA variants outperform trial and error searches; see Sec.~\ref{subsec:GA_comparison}.
  \item (iii) In Sec.~\ref{subsec:lowfit}, we show that when the seed network is of very low quality quality with no predictive ability whatsoever, the genetic algorithm is able to discover new networks with high accuracy and low complexity. This is important when designing entirely new networks where good hyperparameter values may be unknown.
  \item (iv) In Sec.~\ref{sec:FinalGA}, we showed that when starting from the architecture proposed by George and Huerta~\cite{George2018FirstPaper}, the GA-optimized network has 78\% fewer trainable parameters while obtaining an 11\% increase in accuracy for our test problem. This showcases the GA's ability to refine state-of-the-art convolutional neural networks to simultaneously achieve both more compact networks and higher accuracy. In all of our experiments, we find the GA discovers significantly less complicated networks as compared to the seed network, suggesting it can be used to prune wasteful network structures. 
\end{itemize}

High dimensional hyperparameter optimization is challenging. Based on considerations of the problem, evolutionary algorithms in general, and genetic algorithms in particular, are one possible solution to this problem. Future work should include exploring and comparing to alternative algorithms, such as particle swarm optimization or Bayesian optimization, as well as different forms of the GA fitness score.
Indeed, due to the choice of fitness score, the GA optimizes the network at a fixed threshold of the ranking statistic, $P_{\tt signal}^*$, instead of a fixed false alarm probability. In Sec.~\ref{sec:network_comparisons} we see that at a fixed false alarm probability the GA-optimized network does not have a clear accuracy advantage, although it is significantly more compact. In future work we hope to explore different loss functions or GA fitness scores to directly control and optimize for target FAPs. 
Furthermore, as the complexity of the neural network is high, one might consider designing fitness functions for individual layers of the classifier thereby reducing one high dimensional optimization problem to a handful of lower-dimensional optimization problems. While this approach is computationally attractive, it would require access to layer-specific fitness functions that, at least at present, do not have an obvious choice. However, if such fitness functions can be found (perhaps for specific problems) this would provide for faster optimization.

To facilitate comprehensive studies of the GA's behavior, we have focused on signals from non-spinning binary black hole systems. One important extension of our work is to consider GA optimized networks in the context of signals from spinning and precessing binary black hole systems, which is the more realistic case of interest. Finally, while we have restricted our attention to deep CNN classifiers, genetic algorithm optimization can be applied to any other machine learning setting where hyperparameter values need to be set, including alternative architectures for signal classification or parameter estimation. 

GA-optimized networks should prove useful in a variety of contexts. Most importantly, they provide some assurance that the most accurate, compact networks are being found and used in realistic gravitational-wave searches. GA-optimizations should be especially when exploring new architectures or refining an existing one. For example, if the detector's noise properties or signal model might change, a GA can make automated hyperparameter refinements while the network is retrained. GA optimizations may also be useful when comparing different machine learning algorithms. In such comparisons it is often unclear if the better performing model is genuinely better or its hyperparameters are better optimized; automating the hyperparameter selection will remove some of this ambiguity. Finally, the evolution of hyperparameter values over generations might provide insight into the network by elucidating degeneracies and patterns in network's structure. 

\section{Acknowledgments}
We would like to thank  Prayush Kumar, Jun Li, Caroline Mallary, Eamonn O'Shea, and Matthew Wise for helpful discussions, and Vishal Tiwari for writing scripts used to compute efficiency curves. S.\ E.\ F.\ and D.\ S.\ D.\ are partially supported by NSF grant PHY-1806665 and DMS-1912716. G.K. acknowledges research support from NSF Grants No. PHY-1701284, PHY-2010685 and DMS-1912716. All authors acknowledge research support from ONR/DURIP Grant No.\ N00014181255, which funds the computational resources used in our work. D.\ S.\ D.\ is partially supported by the Massachusetts Space Grant Consortium.

\section*{Conflict of interest}

The authors declare that they have no conflict of interest.

\appendix

\section{Fourier transform and inner product conventions}
\label{app:conventions}

We summarize our conventions, which vary somewhat in the literature. Given a time domain vector, $\mathbf{a}$, the 
discrete version of the Fourier transform of $\mathbf{a}$ evaluated at frequency $f_p = p/T$ is given by
\begin{align} \label{eq:DFT}
\tilde a(f_p) = \tilde{a}[p] =  \Delta t \sum_{n=0}^{N-1} a(t_n) e^{-2 \pi i f_p n \Delta t} =
\Delta t \sum_{n=0}^{N-1} a(t_n) e^{-2 \pi i n \frac{p}{N}} \,,
\end{align}
where $0 \leq p \leq N-1$. Notice that the
zero frequency ($f_p=0$) corresponds to $p = 0$, positive frequencies
($0 < f_p < f_s / 2$) to values in the range $0 < p \leq N/2$, and negative
frequencies ($- f_s / 2 \leq f < 0$) correspond to values in the range $N/2 < p < N$.
This follows from
the usual assumptions that the signal is
both periodic in the observation duration, ${a}(t) = {a}(t \pm T)$,
and compactly supported, $\tilde{a}(f) = 0$ for $|f| \geq f_s / 2$,
where $f_s = 1 / \Delta t$ is the sampling rate and $f_s / 2$
is the Nyquist frequency. Consequently,
the Fourier transformed signal is periodic in $k$ with a period of $N$,
$\tilde{a}(f_k) = \tilde{a}(f_k \pm N \Delta f)$.
The value $p = N/2$ corresponds to
the Fourier transform at the maximum resolvable frequencies, $-f_s/2$ and $f_s/2$,
for a given choice of $\Delta t$.

Given the Fourier transformed data, $\tilde{a}$ and $\tilde{b}$,
the noise-weighted inner product $\langle \cdot , \cdot\rangle$
between $\tilde{a}$ and $\tilde{b}$ is defined as
\begin{align}\label{eq:inner_product}
\langle a , b\rangle = 2 \Delta f \sum_{i=0}^{N-1} \frac{a(f_i) b^*(f_i)}{S_n(f_i)}
\approx 2 \int_{-f_s /2}^{f_s /2} \frac{a(f) b^*(f)}{S_n(f)} df \,.
\end{align}
Notice that by convention 
the inner product is defined with an overall factor of $2$, 
but unlike Eq.~\ref{eqn:discreteip} the full set of positive and negative 
frequencies are used. The continuum limit ($\Delta f \rightarrow 0$) of
the summation makes clear that this is a (discretized) inner product between 
$a(f)$ and $b(f)$ over the domain $|f| \leq f_s /2$. 
Note that because the time-domain signal is real the Fourier transformed signal satisfies
$\tilde{a}^*(f) = \tilde{a}(-f)$. As a result, the inner product expression can be ``folded-over"
\begin{align}
\langle a , b\rangle = 4  \Re \sum_{i=0}^{N/2-1} \frac{a(f_i) b^*(f_i)}{S_n(f_i)} 
\approx 4  \Re \int_{0}^{f_s /2} \frac{a(f) b^*(f)}{S_n(f)} df \,,
\end{align}
which now features an integral over the positive frequencies and shows the inner product to be manifestly real. 
We then arrive at Eq.~\ref{eqn:discreteip}. This motivates the use of the term ``inner product" when discussing 
Eq.~\ref{eqn:discreteip} despite the fact that when taken 
at face value it does not satisfy the usual
properties of an inner product while Eq.~\eqref{eq:inner_product} does.
Finally, some authors set the noise at the 
Nyquist frequency to 0 (see, for example, Ref.~\cite{Allen:2005fk}  discussion after Eq. 7.1.)
frequency.

\section{Derivation of conditional probabilities used in likelihood-ratio test}
\label{app:matched_filter}

A derivation of the standard inner product used in gravitational-wave analyses
can be found in Ref.~\cite{Finn:1992}, which makes use of methods laid out in
Ref.~\cite{Wainstein:1962}. Here, we provide a brief derivation to highlight
some of the assumptions that go into the classical filter.

In the absence of a signal, we assume that the detector is a stochastic process
that outputs Gaussian noise with zero mean. The likelihood that some observed
output $\vdata$ is purely noise is therefore given by a $N-$dimensional
multivariate normal distribution
\begin{equation}
\label{eqn:pnoise}
p(\vdata | \noise) = \frac{\exp\left[
    -\frac{1}{2}\vdata^{\transpose}\vcovmat^{-1} \vdata\right]}
    {\sqrt{(2\pi)^{N} \det \vcovmat}},
\end{equation}
where $\vcovmat$ is the covariance matrix of the noise and $\det \vcovmat$ is
its determinant.

It is also common to assume that the noise is wide-sense stationary and
ergodic. This is generally true on the time scales that a gravitational-wave
from a compact binary merger passes through the sensitive band of the detector
($\sim \max \mathcal{O}(100\,\mathrm{s})$). In that case, $\vcovmat$ is a
real symmetric Toeplitz matrix with elements
\begin{equation*}
\covmat[j, k] = \frac{1}{2} \acf[k-j]
\end{equation*}
where
\begin{equation}
\label{eqn:defacf}
\acf[k] \equiv \lim_{n\rightarrow \infty} \frac{1}{n} \sum_{l=-n}^{n-1} \data[l]\data[l+k]
\end{equation}
is the autocorrelation function of the data.

There is no general, analytic solution for $\vcovmat^{-1}$. However,
if $\acf \rightarrow 0$ in finite time $\tau_{\max}$ and the observation time
$T > 2\tau_{\max}$ (i.e., $\lceil N/2 \rceil > \lceil \tau_{\max}/\Delta t
\rceil$), then $\vcovmat$ is nearly a circulant matrix; it only differs in
the upper-right and lower-left corners. All circulant matrices, regardless of
the values of their elements, have the same eigenvectors~\cite{CIT-006}
\begin{equation}
\label{eqn:eigenvectors}
\eigenvec_p[k] = \frac{1}{\sqrt{N}} e^{-2\pi i k p/N}.
\end{equation}
We make the approximation that $\vcovmat$ is circulant, and use these
eigenvectors to solve the eigenvalue equation, yielding
\begin{equation}
\label{eqn:eigenvalue_sum}
\lambda_p = \frac{1}{2} \Re\left\{ \sum_{l=-N/2}^{N/2-1} \acf[l] e^{-2\pi i p l /N} \right\}.
\end{equation}
(The $\Re$ arises because the covariance is real and symmetric.) The error in
this approximation decreases with increasing observation time; indeed, the
eigenvalues of $\vcovmat$ asymptote to Eq.~\ref{eqn:eigenvalue_sum} as $N
\rightarrow \infty$ \cite{CIT-006}. The autocorrelation function of
ground-based gravitational-wave detectors $\approx 0$ for $\tau >
\mathcal{O}(10\,\mathrm{ms})$. Since the observation time for a gravitational
wave is $>\mathcal{O}(\mathrm{s})$, this approximation is valid in practice.

We recognize Eq.~\ref{eqn:eigenvalue_sum} as $1/\Delta t$ times the real part
of the discrete Fourier transform of $\acf[p]$.\footnote{We use the same
convention for the Fourier transform as in Ref.~\cite{Allen:2004gu}.}
Therefore, via the Wiener-Khinchin theorem,
\begin{equation}
\label{eqn:eigenvalue_psd}
\lambda_p = \frac{\psd[p]}{2\Delta t}
\end{equation}
where $\psd[p]$ is the discrete approximation of the power spectral density
(PSD) of the noise at frequency $p/T \equiv p \Delta f$. Since the matrix
of eigenvectors $\veigenmat$ are unitary, we have
\begin{align}
\label{eqn:inverse_cov1}
\covmat^{-1}[j, k] &\approx \left[\veigenmat \vevalmat^{-1} \veigenmat^\dagger\right][j, k] \nonumber \\
                    &\approx \frac{2 \Delta t}{N} \sum_{p=0}^{N-1} \frac{e^{-2\pi i j p/N} e^{2\pi i k p/N}}{\psd[p]} \nonumber \\
                    &= c_{jk} + 4 \Delta f (\Delta t)^2 \sum_{p=1}^{N/2-1} \frac{\cos\left(2\pi(j-k)p/N\right)}{\psd[p]},
\end{align}
To go from the second to the third line, we have substituted $1/N = \Delta f
\Delta t$ and have made use of the fact that $S_n[p]$ is symmetric about $N/2$;
$c_{jk}$ depends only on the $p=0$ and $p=N/2$ terms, which correspond
to the DC and Nyquist frequencies, respectively.

Gravitational-wave detectors have peak sensitivity within a particular
frequency band $[f_0, f_{\max}]$ (for current generation detectors, this is $f
\sim [20, 2000]\,$Hz). Outside of this range we can effectively treat the PSD
as being infinite, making all terms in Eq.~\eqref{eqn:inverse_cov1} with $p <
\lfloor f_0 / \Delta f \rfloor \equiv p_0$ zero. Likewise, if we choose a
sample rate $1/\Delta t > 2 f_{\max}$, then the Nyquist term is also
effectively zero. The exponential term in the likelihood is therefore
\begin{align*}
\left[\vdata^\transpose \vcovmat^{-1} \vdata\right]
    &\approx 4 \Delta f  \sum_{p=p_0}^{N/2-1} (\Delta t)^2 \sum_{j,k=0}^{N-1} \data[j]\data[k]\frac{\cos\left(2\pi(j-k)p/N\right)}{\psd[p]} \\
    &\approx 4 \Delta f \sum_{p=p_0}^{N/2-1} \frac{\left|\tilde{\data}\right|^2[p]}{\psd[p]}.
\end{align*}
In going from the first to the second line we have again recognized the sums
over $j,k$ as the discrete Fourier transforms over the real time-series data.
We can further simplify this by defining the inner product
Eq.~\eqref{eqn:discreteip}, yielding Eq.~\eqref{eqn:noise_likelihood} for the
likelihood.

\section{How to generate Gaussian Noise} \label{app:noise}

Somewhat surprisingly, we are unaware of a resource that describes how to implement 
Eq.~\eqref{eq:noise_fd} to generate time-domain noise realizations. When implementing
this expression one encounters sufficiently many subtleties that we will summarize our
recipe here. 

Eq.~\eqref{eq:noise_fd} specifies the statistical properties satisfied by the
Fourier coefficients of the noise. Note that in the literature similar expressions for the
discrete Fourier transform coefficients are sometimes given, which differs from ours. 

Since the frequency-domain noise, $\tilde{n}(f_i)$, is complex, we need
to be careful when sampling the real and imaginary parts.
For example, if the desired property is $\langle \tilde{n}^*(f_i) \tilde{n}(f_j) \rangle=\delta_{ij}$, then 
\begin{align}
\Re(\tilde{n}(f_i)) \sim {\cal N}(0,\frac{1}{2}) \,, \qquad \Im(\tilde{n}(f_i)) \sim {\cal N}(0,\frac{1}{2})\,,
\end{align}
which gives 
\begin{align}
\langle \tilde{n}^*(f_i) \tilde{n}(f_j) \rangle = \langle \Re(\tilde{n}(f_i))^2 + \Im(\tilde{n}(f_i))^2 \rangle= \frac{1}{2} + \frac{1}{2} = 1 \,.
\end{align}
Furthermore, for real time-domain functions we have $\tilde{n}^*(f) = n(-f)$ and so only the non-negative frequencies are independently sampled. When $f=0$, this condition implies that $n(0)$ is real, whence $\tilde{n}(0) \sim {\cal N}(0,1)$. A similar property holds at the Nyquist frequency.

The neural networks considered in this paper use time-domain data. Synthetic time-domain noise realizations 
are constructed by taking an inverse Fourier transform of our frequency domain noise. 
In the time-domain, Eq.~\eqref{eq:noise_fd} becomes,
\begin{align} \label{eq:noise_td}
\langle n(t_i) \rangle = 0 \,, \qquad \langle n^2(t_i) \rangle  = \frac{\Delta f}{2} \sum_{i=0}^{N-1} S_n(f_i)\,,
\end{align}
which follows directly from Eq.~\eqref{eq:noise_fd} and properties of the Fourier transform. 
We found Eq.~\eqref{eq:noise_td} to be an indispensable sanity test of our time-domain noise realizations.


\bibliographystyle{ieeetr} 
\bibliography{References}


%
%

\end{document}